\documentstyle[12pt]{article}
\input epsf
\newdimen\td \newdimen\tda

\newcount\tc \newcount\tca \newcount\tcb
\newbox\figbox \newdimen\fighgt \newdimen\figwd
\td=\baselineskip \global\baselineskip=\td
\td=\parskip \global\parskip=\td

\newdimen\fighmarg	
\newdimen\figvmarg	
\newcount\nindentadj	
\newdimen\dropadj	
\newdimen\sideadj	
\newdimen\fontht \newdimen\fontdp  
\setbox\figbox=\hbox{A} \fontht=\ht\figbox
\setbox\figbox=\hbox{y} \fontdp=\dp\figbox
\newbox\fontstrutbox
\setbox\fontstrutbox=\hbox{\vrule height \fontht depth \fontdp width 0pt}
\def\fontstrut{%
 \relax\ifmmode\copy\fontstrutbox\else\unhcopy\fontstrutbox\fi}

\def\getfig#1{
 \setbox\figbox=\vbox{\fig{#1}}
 \fighgt\ht\figbox\advance\fighgt\dp\figbox
 \figwd\wd\figbox}

\newif\iffigleft
\newdimen\isize			
\newdimen\lsize			
\newtoks\aboveArgs		
\newtoks\sideArgs		
\def\makeAboveArgs#1{\begingroup  
 \global\aboveArgs={} \tc=#1
 \loop \ifnum\tc>0
  \global\aboveArgs=\expandafter{\the\aboveArgs 0in\hsize}
  \advance\tc by-1
 \repeat\endgroup}
\def\makeSideArgs#1#2#3{\begingroup
 \global\sideArgs={} \tc=#1
 \loop \ifnum\tc>0
  \global\sideArgs=\expandafter{\the\sideArgs #2 #3}
  \advance\tc by-1
 \repeat\endgroup}

\def\ifempty#1{\def\next{#1}\ifx\next\empty}  

\def\sideFig#1#2#3#4{
 \ifempty{#2} \tca=0
 \else \tca=#2 \fi			
 \ifempty{#4} \tda=0in
 \else \tda=\parskip\multiply\tda#4 \fi	
 \getfig{#1}
 \advance\tda-\fighgt			
 \isize=\figwd\advance\isize\fighmarg
 \lsize=\hsize\advance\lsize-\isize
 \makeAboveArgs{\tca}
 \ifcase#3	
  \advance\tda-2\figvmarg \advance\tda-\fontht \advance\tda-\fontdp
 \or		
  \advance\tda-\figvmarg     
 \or		
  \advance\tda-\figvmarg \advance\tda-\fontdp
 \or		
  \advance\tda\fontht
 \fi

 \tcb=0
 \loop \ifdim\tda<0in \advance\tcb 1 \advance\tda\baselineskip
 \repeat
 \ifcase#3	
  \advance\tcb-1
  \divide\tda 2 \advance\tda#2\baselineskip \advance\tda-\baselineskip
  \advance\tda\fontdp \advance\tda\figvmarg
 \or		
  \tda=0in \advance\tda-\fontht
 \or		
  \advance\tda#2\baselineskip \advance\tda-\baselineskip
  \advance\tda\fontdp \advance\tda\figvmarg
 \or		
  \advance\tcb 1
  \divide\tda 2 \advance\tda-\fontht
 \fi
 \advance\tcb\nindentadj
 \iffigleft \makeSideArgs{\tcb}{\isize}{\lsize}
 \else \makeSideArgs{\tcb}{0in}{\lsize} \fi
 \advance\tca\tcb \advance\tca 1  
 \advance\tda\dropadj
 \noindent\hbox{\fontstrut}
 \vbox{\vskip\tda		
 \iffigleft \hbox{\hglue\sideadj\box\figbox}
 \else \hfill\hbox{\box\figbox\hglue-\sideadj} \fi
 \vskip-\tda \vskip-\fighgt}	
 \vskip-\baselineskip \vskip-\parskip
 \parshape\tca\the\aboveArgs\the\sideArgs 0in\hsize
 \global\dropadj=0in \global\sideadj=0in \global\nindentadj=0
}
\def\sideSpace#1#2#3#4#5{
 \ifempty{#3} \tca=0
 \else \tca=#3 \fi			
 \ifempty{#5} \tda=0in
 \else \tda=\parskip\multiply\tda#5 \fi	
 \figwd=#1 \fighgt=#2
 \advance\tda-\fighgt			
 \isize=\figwd\advance\isize\fighmarg
 \lsize=\hsize\advance\lsize-\isize
 \makeAboveArgs{\tca}
 \ifcase#4	
  \advance\tda-2\figvmarg \advance\tda-\fontht \advance\tda-\fontdp
 \or		
  \advance\tda-\figvmarg     
 \or		
  \advance\tda-\figvmarg \advance\tda-\fontdp
 \or		
  \advance\tda\fontht
 \fi
 \tcb=0
 \loop \ifdim\tda<0in \advance\tcb 1 \advance\tda\baselineskip
 \repeat
 \ifcase#4	
  \advance\tcb-1
  \divide\tda 2 \advance\tda#3\baselineskip \advance\tda-\baselineskip
  \advance\tda\fontdp \advance\tda\figvmarg
 \or		
  \tda=0in \advance\tda-\fontht
 \or		
  \advance\tda#3\baselineskip \advance\tda-\baselineskip
  \advance\tda\fontdp \advance\tda\figvmarg
 \or		
  \advance\tcb 1
  \divide\tda 2 \advance\tda-\fontht
 \fi
 \advance\tcb\nindentadj
 \iffigleft \makeSideArgs{\tcb}{\isize}{\lsize}
 \else \makeSideArgs{\tcb}{0in}{\lsize} \fi
 \advance\tca\tcb \advance\tca 1  
 \parshape\tca\the\aboveArgs\the\sideArgs 0in\hsize
 \global\nindentadj=0
}

\def\figNameWrap#1#2{ \def\figprefix{#1}\def\figsuffix{#2} }
\def\fig#1{ \epsffile{\figprefix#1\figsuffix} }

\def\rightSubFig#1#2#3{ \figleftfalse\sideFig{#1}{#2}{0}{#3} }

\def\raiseFig#1{ \dropadj=-#1 }
\def\moveFigLeft#1{ \sideadj=-#1 }

\figNameWrap{}{.eps}

\def\mf{master field\, }

\def\tr{{\rm tr\,}}
\def\Tr{{\rm Tr \, }}
\def\ln{large $N$\,}
\def\1{$1/N$\,}
\def\N{$N= \infty$\,}
\def\oh{{ \textstyle 1 \over 2}}
\def\ad{a^{\dagger}}

\def\re#1{(\ref{#1})}
\def\H{Hermitian\, }
\def\vev #1{\left\langle \Omega| #1 |\Omega\right\rangle}
\newcommand{\bge}{\begin{equation}}
\newcommand{\ee}{\end{equation}}

\begin{document}
\begin{titlepage}
\begin{center}

October, 1994    \hfill    PUPT-1520  \hskip .8truein
\vskip .5in

{{\Large   \bf MASTERING THE MASTER FIELD\footnote{This work was supported in
part   by the National Science Foundation under
grant PHY90-21984.}
}}

\vskip .15in
\vskip .15in

Rajesh Gopakumar\footnote{\tt rgk@puhep1.princeton.edu} \\[.1in]
and\\[.1in]
David J. Gross\footnote{\tt gross@puhep1.princeton.edu}\\[.15in]
{\small {\em
 Joseph Henry Laboratories\\
Princeton University \\
     Princeton, New Jersey 08544 \\}}

\begin{abstract}

The basic  concepts of non-commutative probability theory are reviewed and
applied to the large $N$ limit of matrix models.
We argue that this
is the appropriate framework for constructing the master field
in terms of which   large $N$ theories  can be written.
We explicitly construct the master field in a number of cases including
 QCD$_2$. There we both
give an explicit construction of the master gauge field
and construct
master loop operators as well.
Most important we
extend these techniques to deal with the
general matrix model, in which the matrices
do not have independent distributions and are coupled.
We can thus construct the master field for any matrix model,
in a well defined Hilbert space,
generated by a collection of creation and annihilation
operators---one for each matrix variable---satisfying the Cuntz algebra.
We also discuss the equations of motion obeyed by the master field.

\end{abstract}
\end{center}
\end{titlepage}
\newpage
\section {Introduction}
 \setcounter{equation}{0}

Theories invariant under  $U(N)$  (or $O(N)$)\footnote{In the large $N$ limit
there is no difference between $SU(N)$(or $ SO(N)$)  and $U(N)$ (or $
O(N)$).},  in which the basic dynamical variables are $N^2$
dimensional  matrices in the adjoint representation of the group, simplify
greatly in the limit of  large $N$. In some cases the simplification is so
great
that the $N=\infty$ theory is solvable.
Large $N$ matrix models  are of   great interest for many reasons
\cite{largeN}. First,  QCD  is such a theory, if we regard the number of colors
as a free parameter. There is much evidence that the large $N$ expansion of
QCD  correctly captures the essence of confinement and asymptotic freedom
and that $1/3^2$ is a good expansion parameter. Second, matrix models have
proved useful as
devices for constructing  string theories. Thus the control of the large $N$
expansion of simple matrix models led some years ago to the non-perturbative
solution of toy string models in dimensions less than or equal to two
\cite{2d}. In
fact, QCD itself might be such an example, the large $N$ expansion of QCD might
be described by a string theory, a goal which  has been realized in two
dimensions \cite{g}. Therefore it is important to explore and develop all
available methods for controlling large $N$ matrix models.

One of the most  appealing ideas to emerge in the study of the large $N$  is
that of the {\em master field} \cite{witten}. The idea is that  there exists a
particular classical matrix field such that the large $N$ limit of all $ U(N)$
invariant  Green's functions are given by their values at the master field.
Thus the master field is analogous to the classical field, in terms of which
all correlation functions are determined in the classical, $\hbar \to 0$,
limit; $1/N^2$ playing the role
of $\hbar$. The argument for the existence of such a \mf is simple.
Consider a general matrix model. By this we mean a theory in which the
dynamical variables are $N\times N$ dimensional \H matrices $M_i$ with an
action $S[M_i]$ that is invariant under
the global $U(N)$ transformation
$M_i \to U M_i U^{\dagger}$.
Consider the correlation functions of $U(N)$ invariant observables.  If we have
a  denumerable set of variables then the most general invariant is a  function
of the  traces of  products of the matrices,
i.e., $O={1\over N} \Tr [M_{i_1} M_{i_2}\dots M_{i_n}] $ (normalized so as to
have a finite limit as $N\to \infty$.) For a field theory with continuum
fields, such as QCD, we  also consider continuous products of the matrix
fields, such as Wilson loops.
Denote the expectation value of  $O$ as
\begin{equation}
\langle  O \rangle  \equiv Z^{-1}
\int \prod_i d M_i e^{-S[M_i]} {1\over N} \Tr [M_{i_1} M_{i_2}\dots M_{i_n}] ,
\end{equation}
where $Z=\int \prod_i d M_i \exp(-S[M_i]) $.   The important property of  the
large $N$ limit is that the expectation value of a product of such invariant
observables factorizes\cite{witten,migdal,migmak}:
\begin{equation}
\langle  O_1 O_2 \rangle =\langle  O_1 \rangle\langle  O_2  \rangle + O(1/N^2).
\end{equation}
 This can be proved in perturbation theory by the analysis of the Feynman
graphs. In lattice  QCD it can also be proved order by order in the strong
coupling expansion.
 Consequently, the variance of any invariant observable vanishes in the large
$N$ limit, namely the probability that $O$ differs from its expectation value
is of order $1/N^2$
\begin{equation}
\langle  (O-\langle  O \rangle)^2  \rangle  =\langle  O^2 \rangle -(\langle  O
\rangle)^2 = O(1/N^2).
\end{equation}
This must mean  that the path integral measure is localized on a particular set
of matrices--- the master field---up to a $U(N)$ transformation; just as in the
classical limit the path integral measure is localized, infinitely sharply as
$\hbar \to 0$, on the classical solution of the field equation.

Given the master field, i.e., a  set of $\lq \lq \infty\times \infty "$
matrices $\bar M_i$, all the correlation functions of the invariant observables
are then calculable as
\begin{equation}
\langle  O \rangle  = \tr [\bar M_{i_1} \bar M_{i_2}\dots \bar M_{i_n}] ,
\end{equation}
where no functional integral need be done, we simply evaluate the trace of the
product of master fields.

In a gauge theory, in addition to  the global $U(N)$ symmetry that we used
above we have  a local  $U(N)$ gauge symmetry. In that case, when considering
gauge invariant  Green's functions,  we can only conclude    that the path
integral is localized as $N\to \infty$
on a  single gauge orbit of the gauge group. In other words, if   $\bar
A_\mu(x)$ is the master gauge field then an equivalent master field is $\bar
A^{U }_\mu(x)= U(x)A_\mu (x)U^{\dagger} (x) -i U(x)\partial_\mu U^{\dagger}
(x)$.

Given the master field for a pure gauge theory, say QCD in four dimensions, one
could then calculate the meson spectrum very directly. Since in the \ln limit
the quarks play no dynamical role, quark loops being suppressed by \1, the
quarks are spectators and can be integrated out. Thus, for example, the meson
propagator $G(x,y)=\langle \bar \Psi(x) \Gamma \Psi(x) \bar \Psi(y) \Gamma
\Psi(y)\rangle $, where $\Gamma$ is a matrix in flavor space, is given by
\begin{eqnarray}
G(x,y)&=& \int {\cal D}( \bar \Psi \Psi A_\mu )e^{ -\int d^4x  \left[ \bar \Psi
(i {\not   \partial} -  {\not \! A}- m ) \Psi + {\textstyle{N\over
4g^2}}\Tr(F_{\mu  \nu}F^{\mu  \nu})   \right]} \Psi(x) \Gamma \Psi(x) \bar
\Psi(y) \Gamma \Psi(y)
\nonumber \\ &= &\tr \langle x|{\Gamma}
{1\over i {\not \!  \partial} -  {\not \!\! \bar A}- m}   {\Gamma}{1\over i
{\not \!  \partial} -  {\not \!\! \bar A}- m} |y \rangle .
\end{eqnarray}

Thus, if we knew the master field, $\bar A_\mu(x)$, we could calculate the
meson
spectrum for $N=\infty $. In a gauge theory we can argue further that the
master field can be chosen, by a choice of gauge, to be independent of space
and time! This   is reasonable if we think of the master field as the field
configuration that yields the \ln saddlepoint of the path integral. Since the
action and measure are translationally invariant we might expect that the
saddlepoint is translationally invariant, so that $\bar A_\mu(x)$ and $\bar
A_\mu(0)$ are equivalent up to a similarity transformation,
\begin{equation}
\bar A_\mu(x) = e^{i P  \cdot x}\bar A_\mu(0) e^{-i P  \cdot x},
\end{equation}
where $P_\mu$   plays  the role of the momentum operator. If so,  then we can
perform a gauge transformation $\bar A_\mu(x)\to \bar A^{U }_\mu(x)= U(x)A_\mu
(x)U^{\dagger} (x) -i U(x)\partial_\mu U^{\dagger} (x)$, with $U(x)=
\exp(iP\cdot x)$, to derive a gauge equivalent,
$x_\mu$-independent master field
\begin{equation}
\bar A_\mu = \bar A_\mu(0) +P_\mu, \ \ \bar F_{\mu \nu }= [\bar A_\mu, \bar
A_\nu] .
\end{equation}
Thus the complete solution of \ln QCD would be determined if we could write
down four
 $\lq \lq \infty\times \infty "$ matrices  $\bar A_\mu$!

But what kind of  matrix  is $\bar A_\mu $? What does  an  $\lq \lq
\infty\times \infty "$ matrix mean? What is $U(\infty)$? To make these
questions sharper let us consider
a solvable example of a large $N$ matrix model, a model of $n$ independent \H
matrices,  where the vacuum-to-vacuum amplitude is given by
\begin{equation}\label{ind}
Z= \int \prod_i {\cal D} M_i e^{-N \Tr V_i(M_i)},
\end{equation}
 where $V_i(M_i)$ is an arbitrary polynomial function of $M_i$.

The case of $n=1$ is the one-matrix-model, which is easily solved for \N . The
invariant observables are class functions of $M$, determined by the
eigenvalues $m_i$, $i =1 \dots N$, which for \N yield a continuous function
$m(x=i/N)$. The matrix integral
can be reduced to an integral over $m_i$, by diagonalizing
$M= \Omega m \Omega^{-1}$ (where $m={\rm diag}(m_1, \dots, m_n) $), and using
the fact that
${\cal D} M =  {\cal D}\Omega \prod_idm_i  \Delta(m_i)^2 $, where ${\cal
D}\Omega $ is the  invariant Haar measure on $U(N)$
and $\Delta(m_i)= \prod_{i\leq j} (m_i- m_j)$. The eigenvalue density, $
\rho(m)={1\over N} {dx \over dm}  $  is determined  for \N by the saddlepoint
equation,
\begin{equation}\label{saddle}
\oh  V'(x)  =  \not \!\! \int {d y \rho(y) \over x-y}.
\end{equation}
For the simplest  Gaussian potential, $V(M)={1\over 2} M^2$ the eigenvalues
have the famous Wigner semi-circular distribution
\begin{equation}
\rho(x) = {1\over 2 \pi} \sqrt{4-x^2}.
\end{equation}

Thus, in the one matrix model we can  say that the master matrix is an
 $  \infty\times \infty $ matrix  with eigenvalues $m_i$, where the $m_i$ are
determined by $\rho(x)$, the solution of ( \ref{saddle}). If we now return to
the $n$-matrix model, since the matrices are independent the eigenvalues of
each are   determined, so that we can say that the master matrices are
\begin{equation}\label{master}
\bar M_i = \Omega_i  m^{(i)} \Omega^{\dagger}_i , \ \ \  {\rm where} \ \
m^{(i)}=  {\rm diag}(m^{(i)}_1, m^{(i)}_2 \dots, m^{(i)}_N),
\end{equation}
and the $\Omega_i$ are undetermined unitary matrices. These master  matrices
are perfectly adequate to calculate  decoupled observables such as
$\langle \tr M_i^p\rangle $, in which the $\Omega_i$'s do not appear. However
the general invariant observable in this theory
is a trace of an arbitrary product of different $M_i$'s, namely $\langle
O_\Gamma \rangle$, where  $\Gamma $ denotes an arbitrary {\em word}, i.e., a
free product of $M_i$'s:
\begin{equation}
    O_\Gamma  = {1\over N} \Tr[ M_{i_1}M_{i_2}\dots M_{i_k}\dots]
\end{equation}
 Here the product does depend on the  $\Omega_i$'s and if we choose any
particular  $\Omega_i$'s in (\ref{master}) we would not get the correct answer.
Of course if we integrate over all values of the  $\Omega_i$'s with  Haar
measure  then we get the right result, however this would not be a master field
description.

There is  a direct, but rather ugly, way of dealing with this problem. Consider
the case  $n=2$, with two independent matrices. Write each $N\times N$
dimensional  master matrix as a block diagonal matrix of $K$  $M \times M
$-dimensional matrices, where
$N=M\cdot K$,
\begin{equation}\label{mastn2}
\bar M_i =\left(  \matrix{ \Omega^{(i)}_1 & & & \cr & . & & \cr & & . & \cr & &
& \Omega^{(i)}_K      }\right)\left(  \matrix{ m^{(i)}  & & & \cr & . & & \cr &
& . & \cr & & & m^{(i)}       }\right)
\left(  \matrix{ \Omega^{\dagger (i)}_1 & & & \cr & . & & \cr & & . & \cr & & &
\Omega^{\dagger (i)}_K      }\right).
\end{equation}
The  $\Omega^{(i)}_j, \ j=1 \dots K$ are specific $M\times M$ unitary matrices
chosen at random from the group (with Haar measure) and $m^{(i)} =  {\rm
diag}(m^{(i)}_1, m^{(i)}_2 \dots, m^{(i)}_M) $, with the $m^{(i)}_j$ determined
(as $  M \to \infty$) by the saddlepoint eigenvalue distribution.  The
expectation value of arbitrary words of $M_1$ and $M_2$
will now be correctly given by the trace of these master matrices when we  take
both $K\to \infty $ and $  M \to \infty$. For example consider
\begin{eqnarray}\label{trace2}
&&{1\over N} \langle \Tr [\,M_1 \, M_2 M_1   M_2   \,]   \rangle {\buildrel ?
\over = }  \lim_{M, K \rightarrow \infty}{
{1\over N} {\rm Tr}_N [ \bar M_1  \bar M_2 \bar M_1  \bar M_2 }]\nonumber \\ &&
= \lim_{M, K \rightarrow \infty} {1\over MK}\sum_{j=1}^K {\rm Tr}_M
[ V_j m^{(1)} V_j ^{\dagger}m^{(2)} V_j m^{(1)} V_j ^{\dagger}m^{(2)}],
\end{eqnarray}
where $V_j =\Omega^{(1)}_j\Omega^{(2)\dagger}_j$ are a set of  unitary matrices
chosen at random from $U(N)$. In the limit of   $K\to \infty $, the average
of the product of the $V $'s,
$\lim_{K\rightarrow \infty} {1\over K } \sum_{j=1}^K (V_j)_{a,b} (V_j
^{\dagger})_{c,d}(V_j)_{e,f} (V_j ^{\dagger})_{g,h}$, is  equal to the integral
of the product over Haar measure, which for $M\to \infty$ is:
\begin{equation}
\int {\cal D}V   (V_j)_{a,b} (V_j ^{\dagger})_{c,d}(V_j)_{e,f} (V_j
^{\dagger})_{g,h}= {1\over M^2} [\delta_{ad}\delta_{bc}\delta_{eh}\delta_{fg} +
\delta_{ah}\delta_{bg}\delta_{ce}\delta_{df}] - {1\over
M^3}\delta_{ad}\delta_{bg}\delta_{eh}\delta_{fc}.
\end{equation}
This is simply the law of large numbers.
Inserting this into (\ref{trace2}) yields
\begin{equation}\label{corfour}
 {1\over N} \langle \Tr [\,M_1 \, M_2 M_1   M_2   \,]   \rangle \nonumber  \!\!
  = \!\! {1\over N^3} \left(\! \Tr[\bar M_1^2]
 \left(\!\Tr[\bar M_2] \right)^2\!\!+ \!
 \left(\!\Tr[\bar M_1] \right)^2 \!\Tr[\bar M_2^2]  -{1\over N }
\left(\!\Tr[\bar M_1] \right)^2\! \left(\!\Tr[\bar M_2] \right)^2 \!
\right)\!\! ,
\end{equation}
which is the correct answer. Thus,  the master matrices $\bar M_i, i=1,2$,
given by (\ref{mastn2}), will yield all invariant Green's functions, i.e.,
arbitrary words made out of $M_1$ and $M_2$.
This   construction can be generalized to the case of an arbitrary number of
independent matrices, at the price of imbedding the diagonal matrices $m^{(i)}$
in larger and larger  block matrices. This is a very awkward construction.  It
indicates however the nature of the $\lq \lq \infty\times \infty "$ matrices
that will be required to represent the master field.

Recently I. Singer \cite{singer} has presented an  abstract existence proof for
the  master field for QCD$_2$ and pointed out the relationship to
  the work of Voiculescu on non-commutative probability theory\cite{Voic}.
Indeed,   Voiculescu's methods   yield  a much more satisfactory framework for
representing   the master field for independent matrix models \cite{doug}.
More important we have been  able to  generalize  these methods to deal with
the most
general matrix model, including QCD in any dimension, thus {\em yielding an
explicit representation of the master field for any and all matrix models.} We
do not mean that all matrix models are solvable, but rather that we can define
a well defined Hilbert space
and a well defined trace operation in which the master field of any matrix
model can be explicitly constructed, if one possesses enough information about
the solution of the theory.  Although this  construction can be viewed as
repackaging  it seems  that the
language  that we shall review and develop is  very  appropriate   to
the \N theory and  might lead to new methods for   constructing the master
field, or equivalently  for  solving the \N theory.

In  Section 2 we discuss the general framework of non-commutative probability
theory developed by Voiculescu, define the notion of {\em free random
variables}
and the construction of an appropriate Hilbert space in which the master fields
of models of independent matrices can be constructed. We explore this
construction for the most general such independent matrix model.
We note that the  generating functional introduced by
Voiculescu
in his construction of the representation of a free random variable
has the interpretation of the generating functional of  planar connected
Green's functions.
We also show that the master field can
be regarded as   the solution of a certain {\em master field equation of
motion}.

In Section 3 we consider the explicit  construction of the master field for
some particular solvable  gauge theories. We first find an alternative,
manifestly \H form of the matrix field for independent \H matrix models. We
reformulate the
master equations of motion in a form that is more useful.  We
then discuss the simplest gauge theory,  the  one-plaquette model,
which undergoes a large-N phase transition as a function of coupling. Here we
will find two master fields, one for each region of coupling.

In Section 4 we  turn from theories  of independent  matrices to the general
case
of  coupled matrices.
Based on our interpretation of the generating functional introduced by
Voiculescu
in his construction of the representation of a free random variable
we give a graphical proof of the construction of the master field for
independent
matrices. This argument can then be extended to deal with more general matrix
models.
We show that the master field for any number of coupled matrices can
be formulated within the same Hilbert space as before and give its explicit
construction.
That is,

{\em  \bf If we can solve a  matrix model  then we can write an explicit
expression for the master field as an operator in a well defined  Hilbert
space, whose structure only depends on the number of  matrix variables.}

Section 5 is devoted to the construction of the master field for  QCD$_2$. Here
we shall give an explicit construction of the master field and show that we can
choose a gauge in which it is spacetime independent.

In Section 6 we discuss an alternate description of QCD$_2$ in terms of loops.
We construct
master loop operators based on the observation that simple loops
corresponded to free random variables and that any loop could be decomposed
into words built out of simple loops. The simple structure of
QCD$_2$ is then a consequence of the fact that these form a multiplicative free
family.
We use these master loop  fields to recover the master gauge field.

Finally, in the last section we shall discuss some  of the many directions of
research that are suggested by this construction.

\section{Non-Commutative Probability Theory}
\setcounter{equation}{0}
\vskip .5truein
 Voiculescu has introduced the concept of free random variables for
non-commutative probability theory, which seems to be the appropriate
mathematical framework
for constructing the master field. We shall start by reviewing this framework,
with no pretense at mathematical rigor. For  more details we refer the reader
to \cite{Voic} .

\subsection{Free Random Variables}

For ordinary commuting  random variables the notion of independence is simple,
namely  the probability measure of the random variables $x_i$ factorizes,
$\mu(x_1, \dots, x_n)=\prod_i \mu(x_i)$. Consequently the expectation value of
products of functions of the $x_i$'s factorize
\begin{eqnarray}
&& \langle f_1(x_1)f_2(x_2)\dots  f_n(x_n) \rangle \equiv \int \mu(x_1, \dots,
x_n)  f_1(x_1)f_2(x_2)\dots  f_n(x_n)\nonumber  \\ &=& \prod_i\int
\mu(x_i)f_i(x_i)=  \langle f_1(x_1) \rangle \langle f_2(x_2) \rangle\dots
\langle f_n(x_n) \rangle .
\end{eqnarray}
For non-commuting random variables this definition is much too strong. There is
a weaker definition, that of {\em free random variables} which is conceptually
analogous to independence, though completely non-commutative.

A non-commutative probability space is called free if
\begin{quote}
The expectation value  of products of  functions of  the non-commuting
variables
$M_i$ vanish if the expectation value of {\em all} the individual functions
vanish:
\begin{equation}\label{free}
 \langle f_1(M_{i_1})f_2(M_{i_2})\dots  f_n(M_{i_n})\rangle =0  \ {\rm if}
\cases {  \langle f_i(M_{i_k})\rangle = 0  & {\rm for all}   k=1, \dots, n-1
\cr
{\rm and} \ i_{k}\neq i_{k+1} &\  k=1, \dots, n}.
\end{equation}
\end{quote}
Note that in the above product the neighboring functions must be of different
random variables.

This is a much weaker condition  than the previous definition where, because of
factorization,  the expectation value vanishes if any of the individual
expectation values vanish.
Nonetheless, it is a powerful restriction on the non-commutative probability
space,
that is sufficient  to express all expectation values of products of different
variables  in terms of the individual expectation values. This is shown by
considering the product
\begin{equation}
 \langle \left[ f_1(M_{i_1})- \langle f_1(M_{i_1}) \rangle \right]  \left[
f_2(M_{i_2})- \langle f_2(M_{i_2}) \rangle \right] \dots  \left[ f_n(M_{i_n})-
\langle f_n(M_{i_n}) \rangle \right]\rangle =0.
\end{equation}
Expanding this product one can express the expectation value of a product of
$n$ functions in terms of expectation values of $n-1$, $n-2 \dots $ functions.
Iterating this procedure one  can express the expectation value in terms of
individual expectation values.

The expectation values of free random variables  are obviously
{\em not} symmetric under the interchange of
different non-commuting variables. However, it is a remarkable fact that
if the variables are free, i.e., \re{free} is satisfied,
then the expectation value is cyclically symmetric.
This can be proved using the same strategy we just employed, namely \re{free}
can be used
to inductively show that if the expectation value of $2,3,\dots n$ variables is
cyclic then it
follows that the same is true for the expectation value of $n+1$ variables.
For details see \cite{Voic}.

The advantage of this definition is that independent matrix models in the limit
of \N
are free non-commuting random variables.   To see this we denote
\begin{equation}
\tr [f_1(M_{i_1}) f_2(M_{i_2})\dots  f_n(M_{i_n})] =\lim_{N\rightarrow
\infty}{1\over N}
 \langle \Tr\left[ f_1(M_{i_1})f_2(M_{i_2})\dots  f_n(M_{i_n})\right]  \rangle,
\end{equation}
where the expectation value is taken with the measure given by (\ref{ind}).
Assume that the individual $f_i$'s have vanishing expectation value, $\tr
[f_k(M_{i_k})]=0, k=1, \dots, n$
and consider the  Feynman diagrams that contribute to the product in
 perturbation theory. A given matrix $M_i$ must be contracted, when we use
Wick's theorem,  with the same $M_i$ appearing in another, non-neighboring,
term.   Contracting two $M_i$'s will split the trace into a product of lower
order traces that, when \N,
factorize. Thus one can prove the above claim inductively.
Of course in this case $\tr $ is manifestly cyclic, as it must be for free
random variables.

We can use the fact that independent matrix models describe free random
variables to disentangle the expectation values of  arbitrary words. Thus,
using the above method we see that
\begin{eqnarray}\label{disen} &&  \tr[M_1M_2M_1M_2 ]  =  2\tr[M_1]\tr[
(M_2)^2M_1 ]+ 2\tr[M_2]\tr[ (M_1)^2M_2 ] -(\tr[M_1])^2 \tr[ (M_2)^2]
\nonumber \\ &&   -(\tr[M_2])^2 \tr[ (M_1)^2] -4 \tr[M_1]
\tr[ (M_2)]\tr[  M_1 M_2 ]
+4 (\tr[M_1])^2(\tr[M_2])^2 \nonumber \\ &&  - (\tr[M_1])^2(\tr[M_2])^2 =
\tr[M_1]\tr[ (M_2)^2 ]+  \tr[M_2]\tr[ (M_1)^2 ] - (\tr[M_1])^2(\tr[M_2])^2,
\end{eqnarray}
which agrees with (\ref{corfour}). Therefore we see that the notion of
free random variables automatically captures the content of Haar measure
for independent
matrix variables in the limit of \N.

\subsection{The Hilbert Space Representation of Free Random Variables }

Given a collection of free random variables,  $\{ M_i\}, i=1, \dots n$, the
correlation functions
$\langle M^{n_1}_{i_i} \dots M^{n_k}_{i_k} \dots \rangle$ are linear
functionals
on the free algebra generated by the $M_i$'s. There exists a very general
mathematical
construction that associates elements of a $C^*$ algebra (with a positive
linear
functional $\phi$ defined on it), with operators on a Hilbert space with a
distinguished
unit vector $|\Omega\rangle $.\footnote{
This is the Gelfand-Naimark-Segal(GNS) construction. See \cite{Simon}}
In the case of matrix models of Hermitian or unitary matrices
there is a natural involution operation---the adjoint, so that we wish to
consider
cases in which the above free algebra is actually a $C^*$ algebra.
States on this Hilbert space
are generated by
\begin{equation}\label{gns}
|M_{i_i} \dots M_{i_n}\rangle \equiv \hat  M_{i_i} \dots \hat  M_{i_n}|\Omega
\rangle ,
\end{equation}
where $\hat M_i$'s are the operators that represent the $M_i$'s.
The inner product on this Hilbert space
is defined via the linear functional $\phi$
\begin{equation}
\langle A|B\rangle = \phi(A^\dagger B),
\end{equation}
where $|A\rangle $ and $|B\rangle $ are states
of the form given in \re{gns}.
In particular
\begin{equation} \langle\Omega | \hat  M_{i_i} \dots \hat  M_{i_n}|\Omega
\rangle = \langle\Omega | M_{i_i} \dots M_{i_n}\rangle= \phi( M_{i_i} \dots
M_{i_n}).  \end{equation}

In the case of matrix models where our linear functionals are expectation
values
with respect to the measure
$\prod_i {\cal D}M_i \exp[ -V_i(M_i)]$, together with the trace,
we recognize that the above apparatus is
the appropriate framework for constructing the master matrix operators.
We see from the GNS construction that the required Hilbert space
is huge---a Fock-like space  consisting of states labeled by arbitrary words
in the $M_i$'s. This is in agreement with our discussion of the master field
above
where we argued that the Hilbert space would have to be very large.

For a one-matrix model-involving the matrix $M$ the space is actually quite
simple
and can be described by states labelled by $ |\Omega \rangle $,
$|M \rangle = \hat M |\Omega \rangle$,  $|M^2 \rangle = \hat M^2 |\Omega
\rangle,  \dots
 |M^n \rangle = \hat M^n |\Omega \rangle $.
However, for a matrix model with $n$ independent matrices $M_i$ the Fock space
of words is
isomorphic to the an arbitrary {\em ordered} tensor product of one matrix
Hilbert spaces.
Note that the order is important since $\hat M_1 \hat M_2\hat M_3|\Omega
\rangle \neq
\hat M_1 \hat M_3\hat M_2|\Omega \rangle $.

An ordinary Fock space of totally symmetric or anti-symmetric states is
generated by
commuting or anti-commuting creation operators
acting on the vacuum. We might try to construct the above Hilbert
space in an analogous fashion, by creation operators $\hat \ad_i$,  for each
$M_i$,
acting on the vacuum $|\Omega \rangle$.
However, since the words are all distinguishable we would have to use creation
operators
with {\em no relations}, i.e., there would be no relation between $\hat \ad_i
\hat \ad_j$
and $\hat \ad_j \hat \ad_i$. This is indeed the case. As shown in \cite{Voic}
the above Hilbert space is  identical to the Fock space constructed by acting
on
a vacuum state with creation operators
$\hat \ad_i$, one for each $M_i$, and that
$\hat M_i$ can be represented  in terms of  $\hat \ad_i$ and its adjoint $\hat
a_i$.
Specifically the Fock space  is spanned by the states
\begin{equation}
 (\hat \ad_{i_1} )^{n_{i_1}}(\hat \ad_{i_2} )^{n_{i_2}} \dots (\hat \ad_{i_k}
)^{n_{i_k}}
|\Omega \rangle ,
\end{equation}
where
\begin{equation}
\hat  a_i |\Omega \rangle = 0, \ \  \hat a_i \hat \ad_j = \delta_{ij}.
\end{equation}
This is not an ordinary Fock space. There are no additional relations between
different $\hat a_i$'s
or different $\hat \ad_i$'s, or even  for $\hat a_j \hat \ad_i$, except for the
one that follows from completeness
\begin{equation}
\sum_i\hat  \ad_i\hat  a_i = 1 - P_\Omega =1- |\Omega \rangle \langle \Omega |.
\end{equation}
In the case of the one-matrix model  this implies that $[\hat a,  \hat \ad
]=P_\Omega$.

This algebra of the $\hat a_i$'s and the  $\hat \ad_i$'s is called the Cuntz
algebra.
It can also be regarded as a deformation of the ordinary
algebra of creation and annihilation operators. Indeed it is the $q=0$
case of the q-deformed algebra
\begin{equation}
\hat  a_i \hat \ad_j - q \hat \ad_j\hat  a_i = \delta_{ij},
\end{equation}
an algebra that interpolates between bosons (for $q=1$) and fermions ($q=-1$).
The  above  space can be regarded as the Fock space we would use to describe
the
states of distinguishable particles,
i.e.,  those satisfying  {\em Boltzmann statistics}.\footnote{
Greenberg has discussed such particles with \lq\lq infinite statistics"
\cite{green}}
Working in such as space is very different from working in ordinary bosonic
Fock spaces. In some sense it is much more difficult, since we must remember
the
order in which the state was constructed. Thus simple operators in ordinary
Fock space can become quite complicated here.
For example the number operator in the case $n=1$ is given by
\begin{equation}
\hat N= : { \hat \ad \hat a \over 1- \hat\ad \hat a }: = \sum_{k=1}^\infty(\hat
\ad)^k \hat a^k,
\end{equation}
and obeys the usual commutation relations with $a$ and with $\ad$.
The reason that even such a simple operator is of infinite order in
$\hat a$ and $\hat \ad$ is that it must measure the presence of each particle
in the state, thus it must be the sum of the operators $(\hat \ad)^k \hat a^k$
that count whether a state has a particle in the k$^{\rm th}$
position.
In the general case, for any $n$, the corresponding number operator is given by
\begin{equation}
\hat N= \sum_{k=1}^\infty  \sum_{i_1, \dots i_k}\hat  \ad_{i_1}\dots\hat
\ad_{i_k}\hat a_{i_k}\dots \hat  a_{i_1}  .
\end{equation}
Clearly we need to develop methods for working in such strange spaces.

\subsection{The  Fock Space Representation of $\hat M_i$}

It remains to show that we can construct an operator $M_i$, in terms
of $\hat a_i $ and  $\hat a_i^{\dagger}$ that  reproduces the moments of the
matrix $M_i$.
Thus, suppressing the indices $i$, we  wish to find an operator  $M(\hat a,\hat
\ad)$
in the Fock space so that
\begin{equation}\label{moments}
\tr[M^p]=
\lim_{N\rightarrow \infty}\int {\cal D} M e^{-N\Tr V(M)} {1\over N}\Tr[M^p] =
\vev{\hat M(\hat a,\hat  \ad)}.
\end{equation}
Such an operator is clearly not unique, since  we can always make a similarity
transformation $M\to S^{-1} MS $,  where $S$ leaves the vacuum unchanged
$S|\Omega \rangle=|\Omega \rangle$ and $\langle \Omega|S^{-1} =\langle \Omega|
$.

Voiculescu shows that we can always find such an operator in the form
\begin{equation}
\hat M(a, \ad) = a + \sum_{i=0}^\infty M_n  a^{\dagger n},
\end{equation}
with an appropriate choice of the coefficients $M_n$.
To determine the coefficients   we note that
\begin{eqnarray}
\tr[M ]&=&\vev{\hat M}= M_0; \ \ \  \tr[M^2 ]=\vev{\hat M^2}= M_1+
M_0^2;\nonumber \\  \tr[M^p  ]&=& \vev{\hat M^p}= M_p
+ ({\rm polynomial \ in}  \ M_0, M_1, \dots , M_{p-1}).
\end{eqnarray}
Therefore we can recursively construct $M_0$, $M_1, \dots, M_p$ in terms of
$\tr[M ]$, $\tr[M^2 ], \dots, \tr[M^p ]$.
To  construct the explicit form of these  coefficients we establish the
following lemma.

\noindent {\bf Lemma}
Given an operator of the form $\hat T=\hat a + \sum_{i=0}^\infty t_n  \hat
a^{\dagger n}
$ we associate
the  holomorphic function $K={1\over z} +\sum_{i=0}^\infty t_n z^n $.  Then
\begin{equation}\label{formula}
\vev{F'(\hat T)}= \oint_C {dz\over 2\pi i} F[K(z)],
\end{equation}
where $C$ is a contour in the complex $z$ plane around the origin.

To prove the lemma it is sufficient to prove it for monomial $F$'s, namely to
prove that
 $n\vev{  \hat T^{n-1}}= \oint_C {dz\over 2\pi i}  K^n(z)$. But
$n\vev{  \hat T^{n-1}}= n \Tr[ \hat T^{n-1 }P_\Omega]$. Then we use  the fact
that
$[\hat T, \hat \ad] =[\hat a,\hat \ad]=  P_\Omega $ to write
\begin{equation}\label{eq210}
n\vev{  \hat T^{n-1}}=  n \Tr[  \hat T^{n-1 }[\hat T, \hat \ad]] = \Tr[\hat
T^n,
 \hat \ad],
\end{equation}
where the last equality follows from the fact that $\Tr[\hat T^n, \hat \ad]
= \sum_{i=0}^{n-1} \hat T^i[\hat T, \hat \ad]\hat T^{n-i-1}$ and the fact that
$[\hat T^n, \hat \ad]$
is a trace class operator. Finally we use the fact that if $\hat T_f$ is the
operator associated with the function $f(z)$, that has the Laurent expansion
$f=\sum_{n=-\infty}^\infty f_n z^n$, i.e., $\hat T_f= \sum^{\infty}_{n=1}
f_{-n}\hat  a^n+f_0 +\sum^{\infty}_{n=1} f_n \hat a^{\dagger n}$,  then
\begin{equation}\label{integ}
\Tr[ \hat T_f, \hat T_g] = \oint_C {dz\over 2\pi i} f(z)g'(z).
\end{equation}
It is sufficient to establish this formula  for the case where $f(z)$ and
$g(z)$ are monomials,   then (\ref{integ}) follows by additivity. Consider
  $f(z)=f_nz^n$ so that $\hat T_f=f_n\hat a^{\dagger n}$.  Clearly   $\Tr[ T_f,
T_g]$
will vanish unless $\hat T_g=g_{-n}\hat a^{  n}$, i.e., $g(z)=g_{-n}z^{-n}$.
Using
$\hat a^{\dagger n} |m \rangle=\hat a^{\dagger (n+m)}|\Omega \rangle =|n+m
\rangle$,
\begin{eqnarray}
&& \Tr[ \hat T_{z^n}, \hat T_{z^{-n}}] =  \sum_{m=0}^\infty \langle    m|\hat
a^{\dagger n}\hat a^{
 n}-\hat
a^{  n}\hat a^{\dagger n}|m \rangle= -\sum_{m=0}^{n-1} \langle     m-n | m-n
\rangle
\nonumber \\ &+& \sum_{m=n}^\infty[\langle     m+n | m+n \rangle -\langle
m-n | m-n \rangle]= -n= \oint_C {dz\over 2\pi i} z^n {dz^{-n}\over dz}.
\end{eqnarray}
Using this formula to evaluate  (\ref{eq210})  we establish (\ref{formula}) for
polynomial functions, namely $n\vev{  \hat T^{n-1}}= \oint_C {dz\over 2\pi i}
K^n(z)$.

We now apply this formula to determine the form of the operator $\hat M$ that
reproduces the moments of the matrix $M$. Assuming that we have found such an
operator, so that \re{moments} holds. Then we can express the resolvent,
$R(\zeta)$, the
generating functional of the moments
\begin{equation}
R(\zeta)\equiv  \sum _{n=0}^\infty\zeta^{-n-1}\tr[M^n]=\tr[ {1\over \zeta- M}]
= \int dx {\rho(x) \over \zeta -x},
 \end{equation}
as
\begin{equation}
R(\zeta) =   \sum _{n=0}^\infty\zeta^{-n-1}\vev{\hat M^n}=\sum _{n=0}^\infty
{1\over n+1}\zeta^{-n-1}  \oint_C {dz\over 2\pi i} M^{n+1}(z) =
-\oint_C {dz\over 2\pi i}
\log [ \zeta -M(z)],
 \end{equation}
where $M(z)=1/z +\sum M_nz^n$. Now changing  variables in the integral,
$M(z)=\lambda, z=M^{-1}(\lambda)= H(\lambda)$ we have
\begin{equation}
R(\zeta) =   -\oint_C {d\lambda \over 2\pi i} H'(\lambda)
\log [ \zeta -\lambda ]= \oint_C {d\lambda \over 2\pi i} { H(\lambda) \over
  \zeta -\lambda}= H (\zeta) .
 \end{equation}
Therefore we find that {\em $M(z)$ is the inverse, with respect to composition,
of the resolvent, i.e., $ R(M(z))=M(R(z))=z$}.

This allows us to construct the master field for the one-matrix model
explicitly, since the
resolvent can be constructed algebraically in terms of the potential $V(M)$.
In the simplest case of a Gaussian, $V(M)= {1\over 2\alpha } \Tr[M^2]$, we have
\begin{equation}
 G(z) = {z-\sqrt{z^2 -4\alpha} \over 2\alpha }={2 \over z+\sqrt{z^2 -4\alpha}
} \Rightarrow
M(z) = {1\over z}+\alpha z \, ; \ \ \ \hat M = \hat a + \alpha \hat \ad .
\end{equation}
This form for the Gaussian master field can be made explicitly \H
by a similarity transformation, using the number operator  constructed above.
Indeed if we take $S=\exp[-\oh \log\alpha \hat N]$, then
\begin{equation}
\hat M \to S \hat M S^{-1} = \sqrt\alpha[ \hat a +  \hat \ad] \equiv
\sqrt\alpha \hat x.
\end{equation}

\subsection{ Connected Green's Functions}

 For a non-Gaussian one-matrix model the  master matrix  $\hat M = \hat a+
\sum_{n=
0}^\infty M_n\hat a^{\dagger n}$  will have an infinite number of non-vanishing
$M_n$'s.
The function $M(z)=1/z +\sum M_nz^n$ has, however, a simple interpretation.
Let us recall the relation between the generating functional, $G(j)$,  of
Green's functions and the generating functional of {\em connected Green's
functions },
\begin{equation}
G(j) = \sum_{n=0}^\infty j^n  \langle \tr[M^n]\rangle= {1\over j}R\left({1/
j}\right); \ \  \psi(j) \equiv   \sum_{n=0}^\infty j^n  \langle
\tr[M^n]\rangle_{\rm conn.}=  \sum_{n=0}^\infty j^n \psi^n .
\end{equation}
As shown by Brezin et.al. \cite{brezin} the usual relation that $\psi =
\log[G]$
does not hold for planar graphs.  Rather the full Green's functions can be
obtained in terms of the connected ones by replacing the source $j$ in $\psi(j)
$
by the solution of  the implicit equation
\begin{equation}\label{zeq}
z(j)= j\psi(z(j)).
\end{equation}
Consequently, if one solves \re{zeq} for $z(j)$ then
\begin{equation}
G(j) = \psi (z(j))={1\over j}R\left({1/ j}\right) \Rightarrow R\left({1/
j}\right)= z(j)\Rightarrow
{\psi(z(1/j))\over z(1/j)}= {\psi(R(j))\over R(j)}= j.
\end{equation}
Therefore the the function $\psi(z)/z$ is the inverse, with respect to
convolution, of the
resolvent $R(z)$. But we established above that $M(z)$ is the inverse of
$R(z)$.
Consequently

 {\em \bf The master field function $M(z)$ is such that $zM(z)$
is the generating functional of connected Green's functions.}

This explains why in the Gaussian case $zM(z)=1 +\alpha z^2$, since the only
non-vanishing n-point function is the 2-point function, and why $M(z)$ will be
an infinite series in $z$ for non-Gaussian distributions. Since the resolvent
is a solution of an algebraic equation  of finite order, for a polynomial
potential,
\cite{brezin} it follows that $M(z)$ is a solution of an algebraic equation as
well.
This interpretation suggests a
direct graphical derivation  of the form of the master field that we shall
present in Section 6 and that will prove to be the basis for generalizing this
construction to the case of dependent matrices.

\subsection{ Equations of Motion}

 There are many ways in which independent matrix models can be solved. Saddle
point equations, orthogonal polynomials  or Schwinger-Dyson equations of
motion. The later approach is particularly simple and   leads to equations of
motion for our master fields.
 The Schwinger Dyson equations of motion for  the one-matrix model follow form
the identity
\begin{equation}
 \int {\cal D} M \sum _{ij} {\partial \over \partial M_{ij}}\{ \exp[-N\Tr V(M)]
f(M)_{ij} \}=0,
\end{equation}
for an arbitrary function  $f$ (a sum of polynomials) of $M$.
Using the fact that
\begin{equation}
 {\partial \over \partial M_{ij}}(M^n)_{ab}= \sum_{j=0}^{n-1}
(M^{j})_{ai}(M^{n-j-1})_{jb},
\end{equation}
and the factorization theorem for \N, we derive for $f(M)=M^n$
\begin{equation}\label{schdy}
\langle {1\over N} \Tr[ V'(M) M^n]\rangle =  \sum_{j=0}^{n-1}\langle {1\over N}
\Tr[    M^{j}]    \rangle \langle {1\over N} \Tr[    M^{n-j-1}] \rangle .
\end{equation}
These equations yield recursion relations for the moments of $M$ that can be
used to solve  for the resolvent.

 The  \N equations  can be reformulated in terms of the master field as
\begin{equation}\label{maseq}
\langle \Omega |\left[  V'(\hat M)  -{\delta \over \delta \hat M }
\right]\cdot f(\hat M)|\Omega \rangle =0,
\end{equation}
for arbitrary $f(\hat M)$. In this equation we must define what we mean by the
derivative with respect to the master field. This is defined as
\begin{equation}
{\delta \over \delta \hat M }\cdot f(\hat M)\equiv \lim_{x\rightarrow 0}{f(\hat
M+\epsilon P_\Omega)- f(\hat M) \over \epsilon},
\end{equation}
so that
\begin{equation}
\langle \Omega | {\delta \over \delta \hat M }\cdot \hat M^n|\Omega \rangle=
\sum_{j=0}^{n-1}\langle \Omega |\hat M^{j}  |\Omega \rangle \langle \Omega |
\hat    M^{n-j-1}|\Omega \rangle.
\end{equation}
With this definition \re{maseq} is equivalent to  \re{schdy}.
Below we shall recast these equations in a form that might prove more useful.

\subsection{The Hopf equation}
The Hopf equation appears often in the treatment of large $N$ matrix models.
It arises in the collective field theory description of $QCD_2$
\cite{GM,doug}, where it
 determines the evolution of eigenvalue densities. It is also the equation
of motion of  the $c=1$ matrix model \cite{jev} and governs the behavior of the
Itzykson-Zuber integral \cite{Mat}. We shall see that it arises very naturally
in the
context of non-commutative probability theory  for families of free random
variables.

Let us first introduce the concept of an additive  free family. Given two
free random variables
$\hat M_1$ and $\hat M_2$, with distributions $\mu_1$ and $\mu_2$, their sum
$\hat M_1+\hat M_2$ has a distribution $\mu_3$ denoted by $\mu_1 \oplus \mu_2$.
A one parameter family of free random variables,such that $\mu_{t_1} \oplus
\mu_{t_2}
=\mu_{t_1+t_2}$, will be called an additive free family. In ordinary
probabilty theory the distribution of the sum of two random variables is
given by the convolution of the two individual distributions. However
the Fourier transform is additive, i.e., we add the Fourier transforms
of the individual distributions to get the fourier transform of the
sum. The  non-commutative analog of the Fourier transform isthe ${\cal
R}$-transform  that we have already encountered above.
In section 2.3
we represented the free random variable $M$ by the operator,
\begin{equation}
\hat{M}=\hat  a +\sum_{n=0}^{\infty}M_n\hat a^{\dag n},
\end{equation}
with  the associated  series,
\begin{equation}
M(z)=\frac{1}{z} +\sum_{n=0}^{\infty}M_nz^n \equiv \frac{1}{z} +{\cal R}(z)
\end{equation}
Then it is shown in \cite{Voic} that ${\cal R}(z)$ is additive
\footnote{This also enables one to
establish a central limit theorem
for free random variables \cite{Voic}. }
Namely, if
$\hat M_1$ and $\hat M_2$ are two free random variables  with ${\cal
R}$-transforms ${\cal R}_1$ and
${\cal R}_2$ respectively, then $\hat  M_1 +\hat  M_2$  has a distribution
described by
\begin{equation}
\hat{M}=\hat  a +{\cal R}(\hat  \ad),  \ \ {\rm with } \  {\cal R}(z)={\cal
R}_1(z)+
{\cal R}_2(z).
\end{equation}

It immediately follows that for an additive free family, ${\cal R}(z)$
must be linear in $t$. Thus, for example, a free Gaussian additive family
has
\begin{equation}
M(z,t)=\frac{1}{z}+tz ,
\end{equation}
corresponding to the family of  distributions $\int{\cal D}M \exp
[{-\frac{1}{2t}\Tr M^2}]$.
In general, for an additive free family
\begin{equation}\label{genadd}
M(z,t)=\frac{1}{z}+t\varphi(z).
\end{equation}
where $\varphi(z)$ need not be linear in $z$.

Consider the distribution for the free random variable
$\hat{N}(t)=\hat{N}_0+\hat{M}(t)$ where $\hat{N}_0$ is free
with respect to the $\hat{M}$'s which are Gaussian,
but otherwise has some arbitrary distribution.
Due to the additivity of ${\cal R}(z)$,
\begin{equation}
{\cal R}_N(z)={\cal R}_0(z)+tz.
\end{equation}
We shall show that the resolvent $R(\zeta,t)$, which is the inverse of
$N(z,t)=\frac{1}{z}+{\cal R}_N(z)$
obeys the Hopf equation,
\begin{equation}
\frac{\partial R}{\partial t} + R\frac{\partial R}{\partial\zeta }=0.
\end{equation}
To see this note that if
\begin{equation}
\zeta=\frac{1}{z}+{\cal R}_N(z)=\frac{1}{z}+{\cal R}_0(z)+tz
\end{equation}
then,
\begin{eqnarray}
&& R(\zeta,t) = R(\frac{1}{z}+{\cal R}_0(z)+tz,t)=z \
\Rightarrow\  \frac{dR}{dt}\mid _z=0  \nonumber \\
&& \Rightarrow 0= \frac{\partial R}{\partial t}\mid _{\zeta }+
\frac{\partial R}{\partial\zeta }\mid _t \frac{\partial\zeta }{\partial t}\mid
_z \
=\frac{\partial R}{\partial t}+\frac{\partial R}{\partial\zeta }z \ \
=\frac{\partial R}{\partial t}+R\frac{\partial R}{\partial\zeta }.
\end{eqnarray}
This explains the ubiquitous appearence of  the Hopf equation in large $N$
theories. In particular we canunderstand the origin of the Hopf equation in the
$c=1$ matrix model
\cite{doug}.
It is easy to see from this argument that if instead of being Gaussian
${\hat M}(t)$ were some other additive free family, as described by
\re{genadd}, then the equation for the resolvent $R(\zeta,t)$ would be modified
to
\begin{equation}\label{genhop}
\frac{\partial R}{\partial t} +\varphi(R)\frac{\partial R}{\partial\zeta }=0.
\end{equation}
These are   the collective field theory equations for these general families.

We will show in section 6.4 that the Hopf equation also arises in the
case of multiplicative free families.This will explain why it appears
  in $QCD_2$,
where the Gaussian nature of the master
field will  be responsible for its occurence
(though it will not be the resolvent that will obey the
equation.)

\section{The One-Plaquette Model}

The master field representation that we have constructed  for independent
Hermitian matrices is not manifestly Hermitian. However, as we remarked, there
are many equivalent representations of the master field. In this section we
shall derive a manifestly Hermitian
representation of the master field for independent Hermitian matrices and then
apply this
construction of the simplest model of unitary matrices, the one-plaquette model
that exhibits a large-N phase transition \cite{growit}.

\subsection{Hermitian Representation}

We shall now give a prescription, again not unique, to construct   a Hermitian
master  matrix
$\hat M(a,\ad)=\hat M^\dagger(a,\ad)$ that reproduces the moments of   the
one-matrix model of Hermitian matrices. The idea is to express $\hat M $ as a
function of the
\H operator $\hat x\equiv a +\ad$. But $\hat x$ represents the master  field
for a Gaussian matrix model. Thus writing  $\hat M$ in terms of $\hat x $  is
equivalent to
expressing a   matrix  with an arbitrary distribution in terms of one with a
Gaussian distribution. This can be done directly by a change of variables in
the probability measure of $M$.

Write the  moments of the  matrix distribution,   given in terms of the density
of eigenvalues, as
\begin{equation}
\tr[M^n] = \int d\lambda \rho(\lambda) \lambda^n \equiv  \int {dx \over 2 \pi}
\sqrt{4-x^2 } \ \lambda^n(x),
\end{equation}
where  the function $x(\lambda)$ is a solution of the differential equation
$dx/d\lambda= \rho(\lambda)/ \sqrt{4-x^2 } $.
 Therefore if we are given the eigenvalue distribution $\rho(\lambda)$ we can
construct the master field as
\begin{equation}
\hat M= \lambda( a + \ad)=\lambda (\hat x)\equiv M(\hat x) ; \ \  \ {\rm where
} \ \lambda(x) \ {\rm is \ determined  \ by  } \
 {d\lambda(x)\over dx} ={ \sqrt{4-x^2 } \over   \rho(\lambda)}.
\end{equation}
In the case of many independent matrices $M_i$,  we can find the master fields
in \H form as $\hat M_i=\lambda_i(\hat x_i)=\lambda_i(\hat a_i +\hat \ad_i)$,
with
each $\lambda_i $ being determined  separately from the distribution of
eigenvalues of $M_i$.

The master fields in this representation also obey the master equations of
motion discussed above. It is amusing, and perhaps instructive for more
complicated models,
to reformulate these in a way that allows for the construction of the master
field directly using the equations of motion. The equations of motion
(\ref{maseq}) can be rewritten as
\begin{equation}
\vev{V'(M(\hat x))f(M(\hat x)) -[\hat \Pi, f(M(\hat x)) ]  }=0,
 \end{equation}
where $\hat \Pi $ will be defined to be the conjugate operator  to $\hat M$ in
the sense that
\begin{equation}
 [\hat \Pi, \hat M ] =P_\Omega = |\Omega \rangle \langle \Omega|.
 \end{equation}
Note that on the right hand side of the commutator we have the vacuum
projection operator and not the identity.  Since $\hat M$ is \H we can choose
$\Pi$ to be anti-\H.
Thus in the case of the Gaussian potential, where $\hat M= \hat x$, we have
\begin{equation}
  \hat M  =\hat a + \hat \ad, \ \ \hat \Pi =   \hat p \equiv \oh(\hat a - \hat
\ad).
 \end{equation}
With this definition we have that
\begin{equation}
 [\hat \Pi, f(\hat M )] ={\delta \over \delta \hat M}\cdot f(\hat M ).
 \end{equation}
Therefore the equations of motion are equivalent to
\begin{eqnarray}
&&\vev{V'(M(\hat x))f(M(\hat x))- \hat \Pi f(M(\hat x)) + f(M(\hat x)) \hat \Pi
}\nonumber \\
&=& \vev{V'(M(\hat x))f(M(\hat x))-2  \hat \Pi f(M(\hat x)) + f(M(\hat x)) \hat
\Pi }=0 .
\end{eqnarray}
But, since the states $f(M(\hat x))|\Omega \rangle$ span the Fock space as we
let $f$ run over all functions of $M(\hat x)$, these equations are equivalent
to the condition that
\begin{equation}\label{opeq}
  \left[ V'(M(\hat x))- 2 \hat \Pi   \right]  |\Omega \rangle =0.
\end{equation}

This equation can be use to solve for the master field, ie., given the
potential $V(M)$
solve \re{opeq} for a \H operator $\hat M$ in the Fock space where $\hat \Pi $
is conjugate to $\hat M$.  The first step, given an ansatz for $\hat M=M(\hat
x)$
is to derive an explicit representation of  $\hat \Pi $. To do this we first
note that
\begin{equation}
[\hat p, M( \hat x)]= {M( \hat x_l)-M( \hat x_r)\over \hat x_l-\hat x_l },
\end{equation}
where the labels on the $\hat x$ operators means that we are to expand the
fraction in a power series in $\hat x_l$ and $\hat x_r$ and order the operators
so that all the
$\hat x_l$'s are to the left of all the $\hat x_r$'s. Using this notation we
can then write
\begin{equation}\label{Peq}
\hat \Pi = { \hat x_l-\hat x_l  \over M( \hat x_l)-M( \hat x_r) }\hat p.
\end{equation}
 In this expression, when the operators are ordered,
$\hat p$ appears to the right of all the $\hat x_l$'s and to the left of all
the $\hat x_r$'s.

To illustrate how this goes consider the Gaussian case where $V'(M)=M$. Take
$\hat M= g_1\hat x +g_2 \hat x^2 + \cdots $ Then using \re{Peq}
$\hat \Pi= 1/g_1 \hat p-g_2/g_1^2(\hat x \hat p+  \hat p \hat x) +\cdots $
The equation of motion then reads
\begin{equation}
[g_1\hat x +g_2 \hat x^2-2/g_1 \hat p+2g_2/g_1^2(\hat x \hat p+  \hat p \hat x)
+\cdots ]
|\Omega \rangle =[(g_1 -1/g_1 )\hat \ad +()g_2 - 2g_2/g_1^2 (\hat \ad)^2 +g_2
+\cdots] |\Omega \rangle =0.
\end{equation}
Consequently we deduce that $g_1=1, g_2=0, \dots \Rightarrow \hat M = \hat x,
\ \hat \Pi = \hat x$.

\subsection{The One-Plaquette Model}

 The one-plaquette model describes unitary  matrices  $U$ with the
distribution
\begin{equation}
Z= \int {\cal D} U e^{-{N\over \lambda}\Tr[ U + U^{\dagger}]}.
\end{equation}
We shall derive a master field for $U$ in the
manifestly unitary  form $\hat U = \exp[i H(\hat x)]$, where $H(\hat x)$ will
be the master field for the eigenvalues of  $U$,
\begin{equation}
\langle {1\over N} \Tr[U^n] \rangle =\int d \theta \sigma (\theta) e^{i
n\theta}=\vev{e^{inH(\hat x)}}
\end{equation}
The  \N  eigenvalue distribution was determined in \cite{growit} to be
\begin{equation}\label{1pl}
\sigma(\theta) = \cases{ {2\over \pi \lambda } \cos( {\theta \over 2 }) \sqrt{
{ \lambda \over 2} -\sin^2({\theta \over 2}) } & $\lambda \leq 2$ \cr
{1\over 2 \pi}\left( 1 +{1\over 2 \lambda} \cos \theta\right)& $\lambda \geq 2$
}.
\end{equation}
Following the strategy described above we can construct $H$ by the change of
variables $2 \pi \sigma(\theta)d \theta = \sqrt{4-x^2}dx$ and $H(\hat x)=
\theta(\hat x)$.
It immediately follows from \re{1pl} that for weak coupling the master unitary
field is given by
\begin{equation}
\hat U = \exp[ 2 i \sin^{-1}  \sqrt{\lambda \hat x/   8}  ], \ \ {\rm for }\
\lambda \leq 2 .
\end{equation}
The phase transition is visible in the master field, since ${\lambda \hat x/
8}$
is a Gaussian variable, whose means square value exceeds one for $\lambda \geq
2$,
at which point $\hat U$ ceases to be unitary. In the strong coupling phase
the master field is given by
\begin{equation}
\hat U =e^{iH(\hat x)}, \ \ {\rm where} \  \ H(x)+{1\over 2\lambda} \sin H(x)=
\oh x\sqrt{4-x^2} + 2  \sin^{-1} {x\over 2}
\end{equation}
This master field has the remarkable property that
\begin{equation}
\vev{\hat U^n}= \delta_{n,0}+ {1\over \lambda} (\delta_{n,1} +\delta_{n,-1}).
\end{equation}

\section{ The General Matrix Model}

So far we have discussed only independent matrix models where the action can be
written as $S= \sum_i S_i(M_i)$ and there is no coupling between the various
$M_i$'s. We found that the master fields can be constructed in
a Fock space  in terms of creation $\hat a_i$ and annihilation operators $\hat
\ad_i$, one for each degree of freedom, where the only relation satisfied by
these operators is   $\hat a_i  \hat \ad_j=\delta_{ij}$. Now let us consider
the
most general matrix model with coupled matrices, for example QCD in four
dimensions. One might think that it would be necessary to enlarge the Hilbert
space in which the matrices are represented, or to modify its structure. This
is not the case. We show below that we can construct the master  field in the
same space as before, with no new degrees of freedom or relations between the
$\hat a_i$';s and $\hat \ad_i$'s. The only new feature will be that $M_i$ will
be constructed out of all the $\hat a_j$'s and $\hat \ad_j$'s, not just  those
with $j=i$.

Let us go back to the  construction of the master field for independent
matrices and give a graphical proof that
the master field defined by
\begin{equation}\label{mi}
\hat M_i = \hat a_i + \sum_n \psi^{n+1}_i\hat a^{\dagger n},
\end{equation}
where $zM_i(z)=\psi_i(z)= 1+\sum \psi_i^n z^{n }$ is the generating functional
of
connected Green's functions of the matrix $M_i$, i.e., $ \psi_i(z)=
\sum_{n=0}^\infty \langle \tr[M_i^n]\rangle z^n$, yields the correct Green's
functions. Consider the most general
Feynman graph that contributes to
\begin{equation}
\langle {1\over N} \Tr\left[ M_{i_1} M_{i_2} M_{i_3}\dots  M_{i_n}
\right]\rangle.
\end{equation}
The most general contribution to such a Green's function  can be drawn, as in
Fig. 1, in terms of connected Green's functions.  Fig.1  represents  a
contribution to the \N
Green's function $\langle \Tr [ M_i^2 M_2^2 M_1 M_3^2 M_4M_5^3 M_4 ]\rangle $,
where the solid circles represent the connected Green's functions. We are using
the standard double index line notation for the propagators of the matrices.

\centerline{\epsffile{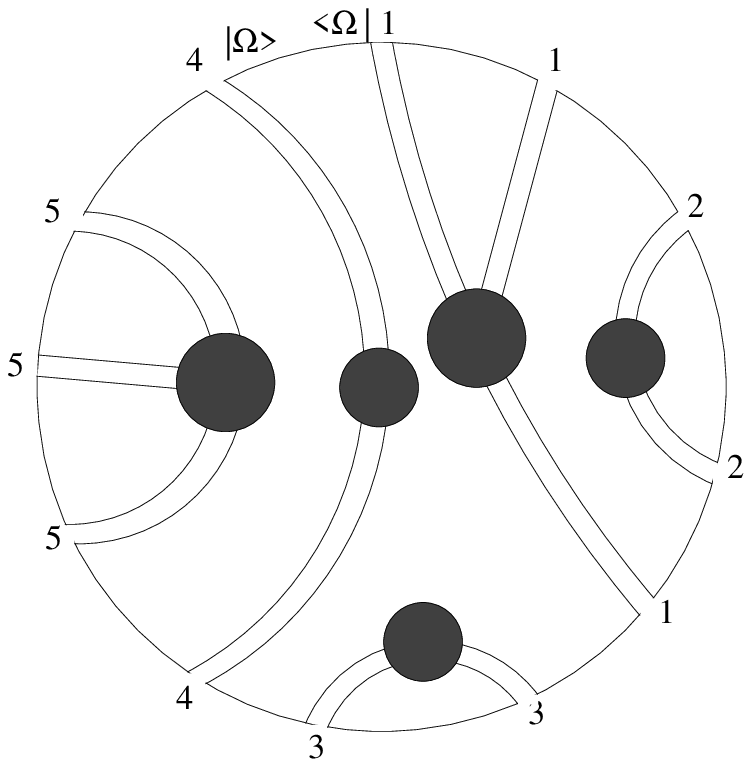}   }
\centerline{ Fig. 1  \hskip .1truein   A  contribution to $\langle \Tr [ M_i^2
M_2^2 M_1 M_3^2 M_4M_5^3 M_4 ]\rangle $. The solid circles represent }
connected Green's functions.

What is special about these graphs is that none of the lines cross, i.e.    the
points around the circle
corresponding to the matrices $M_i$, in the order determined by the above
word,  are joined by lines that do not intersect.  In that case  the double
index graph can be drawn on the plane and contains the maximum number of powers
of $N$.

Now let us note that these graphs are in one-to-one correspondence with the
terms in the expansion of $\vev{\hat  M_i^2\hat  M_2^2 \hat M_1 \hat M_3^2 \hat
M_4\hat M_5^3 \hat M_4} $, with $M_i$ given by \re{mi}.  Writing out the
expression for this vacuum expectation value we find a contribution that
exactly corresponds to the above graph, namely
\begin{eqnarray}
&& \langle \Omega |  ( \hat a_1 + \dots) \cdot (\hat a_1 +\dots )\cdot  (\hat
a_2 +\dots)\cdot  (\dots +  \psi^2_2\hat \ad_2+ \dots)\cdot (\dots +
\psi^3_1\hat a^{\dagger 2}_1+ \dots) \cdot
(\hat a_3 +\dots) \cdot \nonumber \\
&&  \cdot (\dots +  \psi^2_3\hat \ad_3+ \dots) \cdot (\hat a_4 +\dots)\cdot
(\hat a_5 + \dots) \cdot (\hat a_5 +\dots )\cdot (\dots +  \psi^3_5\hat
a^{\dagger 2}_5+ \dots) \cdot \nonumber \\
&& (\dots +  \psi^2_4\hat a^{\dagger}_4+ \dots)| \Omega\rangle = \psi^2_2
\psi^3_1\psi^2_3 \psi^3_5 \psi^2_4  .
\end{eqnarray}
Conversely,   every non-vanishing contribution to  $\vev{\hat  M_i^2\hat  M_2^2
\hat M_1 \hat M_3^2 \hat M_4\hat M_5^3 \hat M_4} $ corresponds to a  planar
graph, such as that depicted in Fig.1. Start along the circle at a point
denoted by $ \langle \Omega |$, then
$\hat M_1$ must create a line labeled by $1$.  The next operator is $\hat M_1$,
so it can annihilate the line $1$, with  coefficient $\psi^2_1$, or
create another $1$ line as in Fig.1. Then the operator $\hat M_2$ cannot
annihilate the line $1$, since $\hat a_1\hat \ad_2=0$.
Then the next $\hat M_2$ must annihilate the line 2, with coefficient
$\psi^2_2$, as in Fig.1. Clearly there is no way we can get graphs with lines
$1$ and $2$ crossed, the Cuntz algebra  ensures that this cannot happen.
So on around the circle.

This argument is a graphical proof that the master fields  is given by \re{mi},
with $\psi_i(z)$ being the generator of connected Green's functions of the
$M_i$'s.
However, this argument does not really depend on the matrices being
independent. Consider a general matrix model, with partition function
\begin{equation}
Z= \int \prod_i  {\cal D} M_i e^{-N \Tr S(M_1,   \dots, M_i) },
\end{equation}
where  $S$ will in general contain interactions between different matrices.
 The general graph that contributes to the Green's function has a similar
decomposition in terms of connected Green's functions. An example of a
contribution to
$\langle \tr\left[ M_{1} M_{2} M_{3}   M_{4}M_{1}M_{3}M_{5}M_{4}
\right]\rangle.$
is depicted in Fig.2
\vskip -.3truein

\centerline{\epsffile{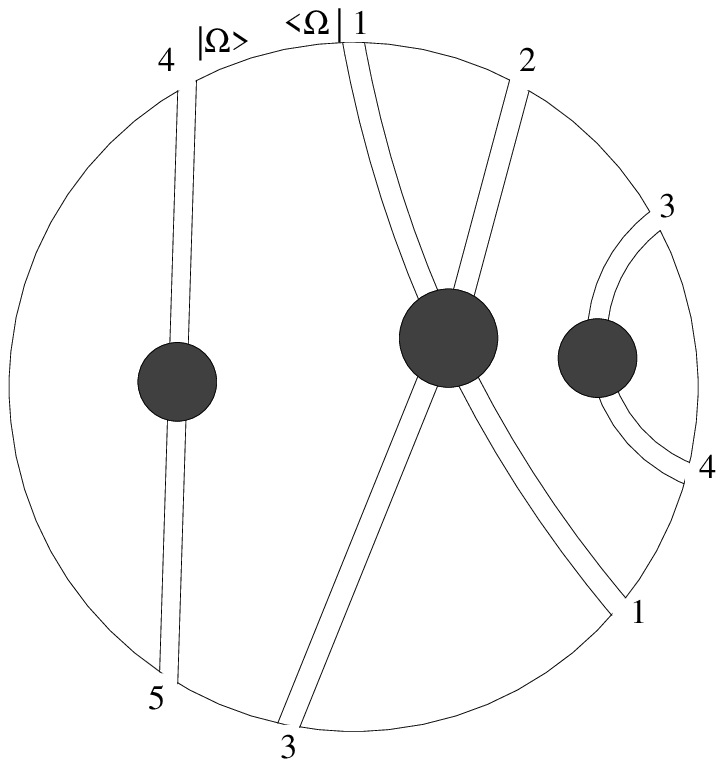}   }
\centerline{ Fig. 2  \hskip .1truein   A  contribution to $\langle\tr [ M_{1}
M_{2} M_{3}   M_{4}M_{1}M_{3}M_{5}M_{4} ]\rangle $. The solid circles represent
}
connected Green's functions.

\noindent The only difference is that the solid circles, representing the
connected Green's functions, now can involve matrices with different indices.
We can  construct a master field for each $M_i$,  in terms of the  same
creation and annihilation operators as before, as long as we let $\hat M_i$
depend on {\em all the creation operators}. Thus if
\begin{equation}
\hat M_i =\hat a_i + \sum_{k=1}^\infty \psi_{i, j_1,j_2, \dots ,j_k}\hat
\ad_{j_1}\hat \ad_{j_2}\dots \hat  \ad_{j_n},
\end{equation}
where  $\psi_{i, j_1,j_2, \dots ,j_k}$ is the connected Green's function:
\begin{equation}
 \psi_{i, j_1,j_2, \dots ,j_k} \equiv {1\over N} \langle \Tr[ M_i M_{j_1}
M_{j_2} \dots M_{j_n}]\rangle_{\rm conn.}
\end{equation}
 then    the vacuum expectation values of products of $\hat M_i$'s
will correctly reproduce the \N limit of the general matrix model. Note that
the coefficients $\psi_{i, j_1,j_2, \dots ,j_k}$ are cyclically symmetric in
the indices.

This  argument shows that the master  field exists as an operator in a
Boltzmannian Fock space constructed with the use of a creation operator for
each independent matrix field
and gives an explicit expression for the master fields in terms of the solution
of the  theory. It is not, of course,  an explicit expression for the master
fields--to do so would be to solve the theory.

One new approach to solving large $N$ theories might be to explore the
equations of motion for the master field operators. In the case of coupled
matrices the Schwinger-Dyson equations can also be formulated as
\begin{equation}
\vev{ \left[{\partial S[\hat M_1,\hat M_2,\dots] \over \partial \hat M_i}-
{\delta \over \delta \hat M_i}\cdot  \right] f[\hat M_1,\hat M_2,\dots] }=0,
\end{equation}
where$ {\delta \over \delta \hat M_i}$ is defined as before and $f$ is an
arbitrary funtion of the $\hat M_i$. We can also define a conjugate operator,
$\hat \Pi_i$, to $\hat M_i$
as before, that satisfies $[ \hat \Pi_i,\hat M_j]=\delta_{ij} P_\Omega$. If we
can find a \H representation of $\hat M_i$ then the above equations of motion
are equivalent to
\begin{equation}
 \left[{\partial S[\hat M_1,\hat M_2,\dots] \over \partial \hat M_i}-
 -2\hat \Pi_i \right] |\Omega \rangle =0.
\end{equation}
It might be fruitful to explore this equation as a way of solving large $N$
theories.

\section{Two Dimensional QCD}

We now turn to discuss the master field for the simplest gauge theory,
two-dimensional QCD. There are two approaches to the master field that one
might pursue. One is to construct a master loop field, $\hat U_C$, that would
be used directly to reproduce the expectation values of the Wilson loops for
\N. We shall  discuss this approach in the following section. The other
approach is to construct directly the master gauge field
$\hat A_\mu(x)$. This is simple for QCD$_2$,  since in an appropriate gauge
the theory is Gaussian and corresponds to an  independent matrix model, albeit
one with continuum labels.

Consider two-dimensional  Yang-Mills theory in flat space. We can work either
in Euclidean  space, $R_2$, or in Minkowski space $M_2$.  If we choose an axial
gauge $n^\mu A_\mu=0$,
where $n^\mu$ is a unit vector in any direction the theory becomes Gaussian.
This is a legal gauge on $R_2$ or  $M_2$.  We choose for convenience the gauge
$A_1(x)=0$
and work in Euclidean space,  in which case
\begin{equation}
F_{10}(x)=F(x)= \partial_1 A_0 ; \ \ {\cal L} =  {\textstyle{1\over 2}} \Tr
F(x)^2=
{\textstyle{1\over 2}} \Tr\left( \partial_1 A_0(x)\right)^2.
\end{equation}
The field strength $F$ is therefore an independent Gaussian matrix at each
point in spacetime and  we can easily construct a master field to represent it
in the
continuum Fock space defined by
\begin{equation}
\hat a(x)\hat  \ad(y) = \delta^2(x-y); \ \ \hat  a(x) |\Omega \rangle =0; \ \
|x_1, x_2, \dots , x_n\rangle =
\hat \ad (x_1)\hat  \ad (x_2)\cdots\hat  \ad (x_n) |\Omega \rangle.
\end{equation}
In this space we can write
\begin{equation}
\hat F(x) = \hat a(x) +\hat \ad(x).
\end{equation}
Alternatively we can work in momentum space, where
$\hat a(p)=\int {d^2x\over 2\pi} \exp(ip\cdot x)\hat a(x)$ satisfies,
$\hat a(p)\hat \ad(q) = \delta^2(p-q)$. We can then solve the equation
$\partial_1 A_0(x) =F(x)$
to construct the master gauge field
\begin{equation}
\hat A_0(x) = {1\over \partial_1}[\hat a(x) +\hat \ad(x)]= \int {d^2p\over 2\pi
}{i\over p_1} \left[ e^{-ip\cdot x}\hat a(p)- e^{ip\cdot x}\hat \ad(p) \right].
\end{equation}

We can now make a gauge transformation to a gauge in which the master gauge
field will   be independent of $x^\mu$. To this end we define the momentum
operator,
$\hat P^\mu$, in Fock space so that
\begin{equation}
\hat P^\mu |p_1, p_2, \dots, p_n\rangle =
\left( \sum_{i=1}^n  p^\mu_i \right)| p_1, p_2, \dots, p_n\rangle.
\end{equation}
An explicit representation of $\hat P^\mu$ in terms of creation and
annihilation operators is
\begin{equation}
\hat P^\mu =\hat P^{\mu \dagger} = \sum_{k=1}^\infty \int d^2p_1 \dots d^2p_k
p^\mu_k \hat \ad(p_1)\cdots \hat \ad(p_k) \hat a(p_k)\cdots \hat a(p_1) .
\end{equation}
Thus when the $k^{\rm th}$term  in the sum acts on an $n$ particle state, for
$k\leq n$, it removes $k$ particles from the state, measures the momentum of
the $k^{\rm th}$
particle and then puts the $k$ particles back in the original order.
The momentum operator has the standard commutation relations with the creation
and
annihilation operators
\begin{equation}
[\hat P^\mu, \hat a(p)] = -p^\mu \hat a(p), \ [\hat P^\mu,\hat \ad(p)] = p^\mu
\hat \ad(p) \Rightarrow
[\hat P^\mu,\hat A_0(x)] =-i \partial^\mu \hat A_0(x).
\end{equation}
Therefore $A_0(x)= \exp(i\hat P\cdot x)A_0(x)\exp(-i\hat P\cdot x)$ and, as
discussed in the introduction, we can make a gauge transformation on the above
master field, with gauge function $\hat U=\exp(i\hat P\cdot x)$, to derive a
spacetime independent master field:
\begin{equation}
\hat A_1 = \hat P_1, \ \ \  \hat A_0 = \hat P_0 + \hat A(0) =  \hat P_0 +  \int
{d^2p\over 2\pi }{i\over p_1} \left[  \hat a(p)-  \hat \ad(p) \right].
\end{equation}
In this gauge the field strength is given by
\begin{equation}
\hat F_{10}=i[ \hat A_1,\hat A_0]= [\hat P_1, \hat A(0)]= \int {d^2p\over 2\pi
}
 \left[  \hat a(p)+\hat \ad(p) \right]=\hat F_{10}(0).
\end{equation}
By means of a similarity transformation,  we can rewrite $\hat A_0$ as
\begin{equation}
\hat A_0 =    \hat P_0 +  \int {d^2p\over 2\pi }  \left[  \hat a(p)+ {1\over
p_1^2} \hat \ad(p) \right],
\end{equation}
in which the second term we recognize as a sum of master fields for a continuum
of Gaussian
matrix variables, where the momentum space connected two-point function, the
propagator, is $1/ p_1^2$. One has to be careful in  using this field to
introduce an infrared regulator for small $p_1$. This can be done by cutting
out a small hole in momentum space or by a principle value prescription for the
propagator. As explained in \cite{thooft,callan} gauge invariant observables do
not depend on the regularization.

This master field  satisfies the master equations $[ D^\mu, \hat F_{\mu
\nu}(x)] = \delta / \delta \hat A_\nu(x)  $. In the original axial gauge this
means that
\begin{equation}
\langle \Omega |[ \partial_1^2 \hat A_0(x) -{ \delta \over \delta \hat
A_0(x)}\cdot ] f(A_0) | \Omega \rangle =0,
\end{equation}
where as before the operator derivative is defined as
\begin{equation}
{ \delta\over  \delta \hat A_0(x)}  f(A_0) =\lim_{\epsilon\rightarrow 0} {
f(A_0(y) + \epsilon \delta(y-x)P_{\Omega}) -f(A_0(y)) \over \epsilon}, \ \ \
P_{\Omega}= |\Omega \rangle\langle \Omega | .
\end{equation}

These equations of motion can be used to show that the Wilson loop, which can
be written in terms of the master field as,
\begin{equation}\label{wilmas}
\langle W_C\rangle = \langle \Omega |T\{\exp[ ig \int_0^1 \hat A^\mu \dot
x_\mu(t)] \} | \Omega \rangle =\langle \Omega |T\{\exp[ ig \int_0^1 \hat
A^0(x(t)) \dot x_0(t)] \} | \Omega \rangle,
\end{equation}
satisfies the Migdal-Makeenko equations \cite{migmak}. Note that in the first
integral, written in terms of the spacetime independent master field the path
ordering is still necessary since, for non-straight paths, $\hat A^\mu \dot
x_\mu(t)$ do not commute for different $t$'s.

\section{Master Loop Fields}
\subsection{Wilson loops}
One can  alternatively describe the master field
for  $QCD$ in terms of Wilson loops. These are manifestly
gauge invariant and contain, in principle, all information about
gauge invariant quantities. They are also the natural variables for
a string theory formulation of $QCD$ . One would
therefore like to have  master loop operators to describe the
large $ N$ limit of these loops.

 From the point of view of the master field of
$QCD$, the loop approach is, in general, quite unwieldy.The space of loops is
too large and overcomplete.
There is  a lot of redundancy in defining master loop operators for
every possible loop. The space of loops is much bigger than the space of
points.  It is hard to see what  a \lq basis' might be  in this space.
Moreover, to extract information about, say,   the
spectrum of  meson bound states seems extremely difficult in practice starting
from these loops (even in $QCD_2$).

Nevertheless, in $QCD_2$, the loop space and loop variables have many nice and
simplifying
features (mainly due to the area preserving diffeomorphism symmetry of the
theory). These features of the  loop variables are not immediately apparent
from
the master connection that we constructed above.
More  importantly,
they enable one to explicitly construct master loop operators that reproduce
an arbitrary loop average fairly easily. Starting from these loops we can, by
considering infinitesimal loops,
  recover the master field of Section 4.  Alternatively, it will also
be possible to start from the master field and derive the master loops
without too much effort.

\subsection{Free Random Variables in the Loop Space of  QCD$_2$}

The main tool in trying to solve for Wilson loop averages are the
Makeenko-Migdal
loop equations \cite{migmak}.
In 2 dimensions, they are especially tractable   and Kazakov and Kostov have
shown
how the   average of an arbitrary  Wilson loop
can be calculated for \N using these equations. We  shall approach the
calculation of
  loop averages in a somewhat different and suggestive manner.
We shall start by decomposing  an arbitrary loop into a {\it word} built of
simple loops,
all originating at some common base point but with otherwise
non-overlapping interiors.
(By a simple loop we shall henceforth mean a non-self intersecting loop on the
plane.)
We shall   argue that these simple loops form a family of
free random variables. This  will enable us to calculate an
arbitrary loop average in terms of $<{\tr} [U_{C_i}^n]>$'s where the
$U_{C_i}$'s
are the free random variables for  simple  loops $C_i$.
Simple loops will thus  form   a basis in loop space, though
they still contain too much information and are overcomplete.

Let's first show that a set of simple loops, based at one point and
non-overlapping, correspond to  free random variables. For concreteness
consider
the loops $C_1,C_2,C_3$ based at the point $P$ as in Fig.3.

\raiseFig{0.6truein}
\rightSubFig{simple}{0}{1}
We denote the
holonomies along the loops $C_i$ by $U_i$, i.e
\begin{equation}
U_i =  P\exp (i\oint_{C_i} A_{\mu}dx^{\mu})
\end{equation}
Note that the $U_i$'s are $U(N)$ matrices. We claim that

\noindent the $U_i$ have
independent distributions and hence
 are free random variables in the large $N$
limit.
One way to see this is to use the heat kernel action, which we know
to be exact in the continuum limit,
\begin{eqnarray}
{\cal Z} =\int \prod_L {\cal D}U_L \prod_{\rm plaquettes} \!\!\!\!Z_P[U_P];  \
\ \ \
Z_P[U_P]= \sum_R d_R\chi_R[U_P] e^{-C_2(R)A_P},
\end{eqnarray}
where  the sum runs over all    reresentations of $U(N)$,
$\chi_R$ is the character of the representation and $C_2(R)$ its second
Casimir operator. We can  choose a triangulation of the plane such that the
given contours,$C_i$,
are the
borders  of some  of the triangles. The self similar nature of the heat kernel
always allows us to choose such a triangulation.
Then, when we come to use this measure to calculate averages of products of
the $C_i$'s,  we can integrate out all the other link variables, (this is
only true on the plane), leaving us with an equivalent measure ${\cal Z} =\int
\prod_i {\cal D}U_i \prod_i Z[U_i]$, where the product runs over all  the
simple loops, $U_i$.
 Therefore the resulting
integrals over the $U_i$ are over independent distributions.

Naturally, this can also be seen directly from the loop equations. We shall
give a
rough sketch below.
The loop equations are the \N Schwinger-Dyson equations for
Wilson loops.  \cite{migmak},
\begin{equation}
\frac{\partial}{\partial x_{\mu}}\frac{\delta}{\delta\sigma_{\mu\nu}}
W(C)\mid _{x=x(\tau)}= \not \!\! \int dx_{\nu}(\sigma)\delta^{(2)}
(x(\sigma)-x(\tau))W(C_1,C_2)
\end{equation}
Here the L.H.S. refers to a variation of the area of the  loop
by $\delta\sigma_{\mu\nu}$ at $x=x(\tau)$. The R.H.S. vanishes
unless $x(\tau)$ is a point of self intersection in which case
$C_1$ and $C_2$ are the two loops into which C breaks up at that
point. (The $\not \! \int$ refers to the exclusion of $\sigma=
\tau$ in the integral.) In the large $N$ limit, $W(C_1,C_2)=W(C_1)W(C_2)$
and,  as shown by  \cite{kazkos}, because of the area preserving symmetry of
$QCD_2$,
the  loop equations
simplify to (See Fig.4)
\centerline{\epsffile{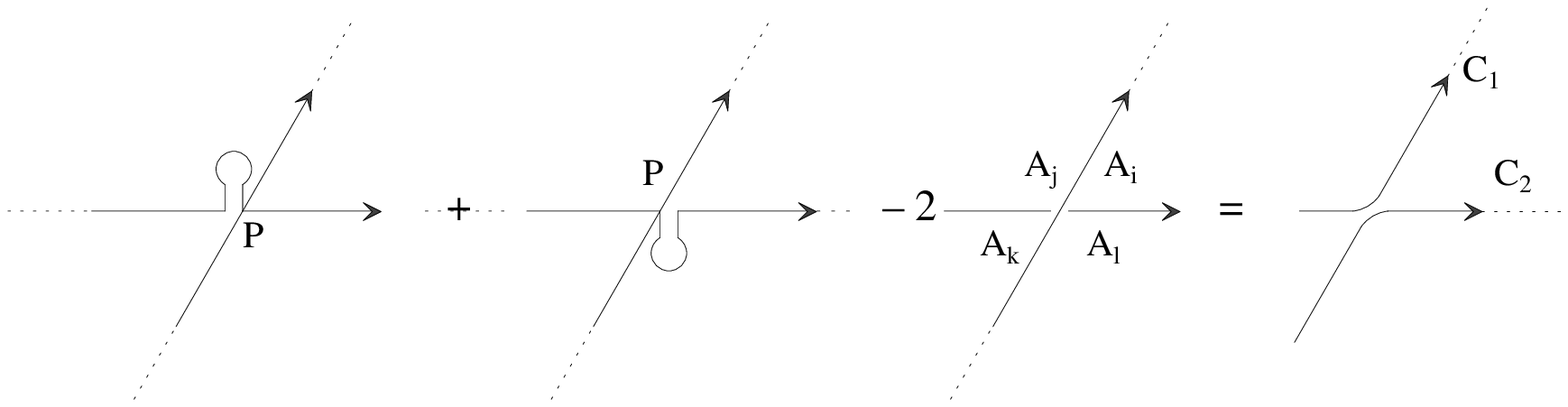}   }
\centerline{ Fig. 4  \hskip .1truein   The Loop Equations for QCD$_2$ }

\begin{equation}\label{loopeq}
(\partial _k + \partial _i - \partial _l -\partial _j)W(C)=W(C_1)W(C_2),
 \ \ \ (\partial _i \equiv \frac{\partial}{\partial A_i}).
\end{equation}
where the $A_i$ are the areas that meet at the point P of self
intersection, at which the loop splits up into $C_1$ and $C_2$. (the R.H.S.
of Fig.4).
The equations (\ref{loopeq}) form  a closed set of equations that  determine
the loop
average for a loop with $n$
self intersections in terms of ones with  a lesser number of self
intersections. They can be solved recursively in terms of the  loop average for
a simple loop.
The latter can be computed  either from perturbation theory or by imposing
appropriate boundary conditions on loops with a  large  number of turns. It has
the value
$W(C)=e^{-\frac{A}{2}}$, where $A$ is the area of the loop.

If we now consider two simple loops, as before, based at some point and with
holonomies $U$ and $V$, then it is possible to compute
$<{\tr} [U^{n_1}V^{m_1}\ldots U^{n_k}V^{m_k}]>$ using the loop
equations. They give a recursion relation for such a word of length $2k$ in
terms of shorter words. We simply state the general expression.
\begin{eqnarray}\label{recur}
<\tr [U^{n_1}V^{m_1}U^{n_2}V^{m_2} \ldots  U^{n_{k-1}}V^{m_{k-1}}
U^{n_k}V^{m_k}]>\nonumber \\
=<\tr [U^{n_1}]><\tr [V^{m_1}\tr U^{n_2}V^{m_2}\ldots  U^{n_{k-1}}V^{m_{k-1}}
U^{n_k}V^{m_k}]> \nonumber  \\
-<\tr [U^{n_1}V^{m_1}]><\tr [  U^{n_2}V^{m_2}\ldots  U^{n_{k-1}}V^{m_{k-1}}
U^{n_k}V^{m_k}]> \nonumber \\
+\ldots -<\tr [ U^{n_1}V^{m_1}U^{n_2}V^{m_2} \ldots
 U^{n_{k-1}}V^{m_{k-1}}]><\tr [U^{n_k}V^{m_k}]> \nonumber \\
+<\tr [ U^{n_1}V^{m_1}U^{n_2}V^{m_2} \ldots  U^{n_{k-1}}V^{m_{k-1}}
U^{n_k}]><\tr [ V^{m_k}]> .
\end{eqnarray}
This recursion relation  can be obtained by combining the loop equations at the
vertices where the loop on the L.H.S. breaks up into the loops represented by
the various terms on the R.H.S.
This relation is easily seen to be  equivalent to a uniform (Haar) distribution
(at large $N$) for
the relative angular integrals between $U$ and $V$. Thus it  implies that
$U$ and $V$ are free random variables.  Conversely, if we were to assume that
the  holonomies of simple loops are  free random variables we could derive this
relation from the defining property of such variables, as  we discussed
previously.

The above  recursion relation is a very useful expression that will enable us
to calculate arbitrary loop averages rather
efficiently.
For $k=1$,
\begin{equation}
<{\tr} [U^{n_1}V^{m_1}]>=<{\tr} [U^{n_1}]><{\tr} [V^{m_1}]>
\end{equation}
and for $k=2$,
\begin{equation}
<{\tr}[UVUV]>=<{\tr}U>^2<{\tr}[V^2]>-<{\tr}U>^2<{\tr}V>^2+<{\tr}[U^2]>
<{\tr}V>^2
\end{equation}
which tallies with  (\ref{disen}).
If we have more than two  such simple loops then these relations can be applied
repeatedly to reduce the average to a product of averages of powers of the
individual $U_i$'s.

The loop equations (\ref{loopeq}) can also be used to compute
$<\tr [U^n]>$ for a simple loop with holonomy $U$. Later we will obtain the
same answer by other means as well.
The answer
can be expressed in terms of Laguerre polynomials
$L^1_n$  \cite{GM,Rossi},
\begin{equation}\label{Lag}
<\tr [U^n]>=<\tr[ U^{-n}]>=<\tr [U^{\dag n}]>  =\frac{1}{n}
L_{(n-1)}^1(nA)e^{-n\frac{A}{2}}=\frac{1}{n}
\oint \frac{dz}{2\pi
i}(1+\frac{1}{z})^n e^{-n\frac{A}{2} (1+2z)},
\end{equation}
where $A$ is the area of  the loop
and  we have exhibited the  integral representation for the  Laguerre
polynomials.
The first few terms  are displayed below.
\begin{eqnarray}
<\tr U>=e^{-\frac{A}{2}},    \ \
<\tr [U^2]>=(1-A)e^{-A},  \ \
<\tr [U^3]>=(1-3A+\frac{3}{2}A^2)e^{-\frac{3A}{2}}.
\end{eqnarray}

\subsection{Decomposing a Loop Into a Word}

Having seen that simple, non overlapping loops, based at a point are free
random variables, we now
proceed to show
how an arbitrary loop can be written as a word built out of such simple loops.
In fact, for a loop with $n$ self intersections, there are $(n+1)$ windows
(i.e. enclosed interiors)
and the word will be built out of $\{U_i\}'s, \ i=1, 2 \dots n+1,$ which will
be associated with these windows.
In the interests of clarity and to avoid notational clutter, we shall
illustrate the decomposition in a few representative cases and its general
nature will  then be apparent.

\raiseFig{0.1truein}
\rightSubFig{fig8}{0}{1}
The simplest self intersecting loop
is the figure of eight in Fig.5.  The loops $C_1$ and $C_2$ are simple loops
and hence $U$ and $V$ are free random variables. Therfore the loop average is
simply
\begin{equation}
W(C)=<{\tr}[ UV]>=<{\tr} U><{\tr}V>=e^{-\frac{(A_1+A_2)}{2}}.
\end{equation}

\eject

\raiseFig{0.3truein}
\rightSubFig{fig6}{1}{1}
The first, non-trivial example is the loop in Fig.6a.
We shall introduce a notation for  loops in terms
of line segments. Thus (13) denotes $a$ and $(\overline{31})$, $b$. The bar is
to
distinguish it from (31) which will denote $a^{-1}$ -- the oppositely
directed line segment to (13) and similarly $(\overline{13})$ for $b^{-1}$.

Then, together with (12)=$c$ and
$(\overline{21})=d$, the loop itself can be written as
\begin{eqnarray}
C= (13)(\overline{31})(12)(\overline{21})
=(13)(\overline{31})(\overline{12})(21)(12)(\overline{21})(12)(\overline{21})
\end{eqnarray}
Where we have inserted (21)(12)=1
and $(\overline{12})(\overline{21})=1$ (two back tracking loops enclosing zero
area).
This is geometrically equivalent to the loop
in Fig.6b. As a word we see that it is $UV^2$ where $U$ corresponds to the loop
$(13)(\overline{31})(\overline{12})(21)$ and $V$ to the loop
$(12)(\overline{21})$.
These  are simple loops and therefore,
their holonomies $U$ and $V$ respectively, have independent distributions.
Therefore,
\begin{equation}
W(C)=<\tr [UV^2]>=<\tr U><\tr [V^2]>= e^{-\frac{A_1}{2}}e^{-A_2}(1-A_2)
\end{equation}
which is the standard answer.

\eject
\centerline{\hskip .4truein \epsffile{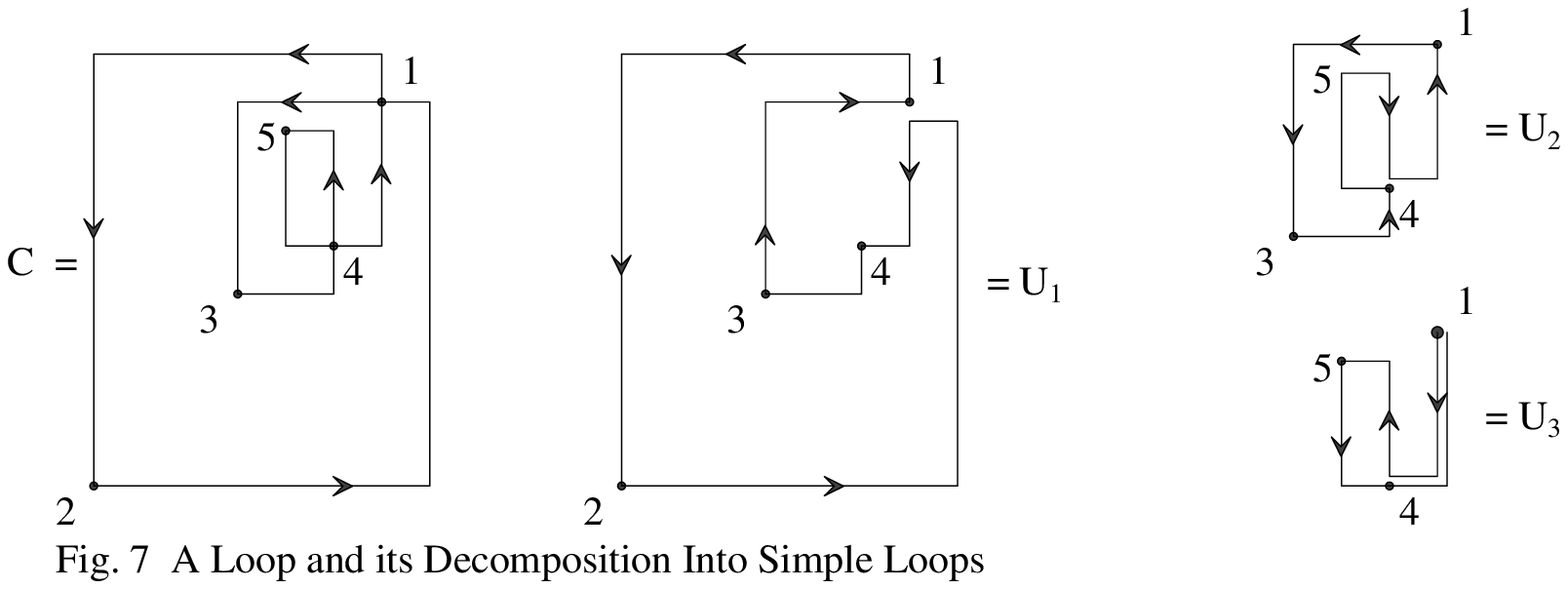}   }
Next consider the contour in Fig.7. With the points labelled as shown,
\begin{eqnarray}
 C&=&(12)(\overline{21})(13)(34)(45)(\overline{54})(41) \nonumber \\
&=&\underbrace{(12)(\overline{21})(14)(43)(31)}_{U_1}(13)(34)(41)(13)(34)(45)
(\overline{54})(41) \nonumber \\
&=&\underbrace{(12)(\overline{21})(14)(43)(31)}_{U_1}\underbrace{
(13)(34)(\overline{45})(54)(41)
}_{U_2}\underbrace{(14)(45)(\overline{54})(41)}_{U_3} \nonumber \\
&&\underbrace{(13)(34)(\overline{45})(54)(41)}_{U_2}
\underbrace{(14)(45)(\overline{54})(41)}_{U_3}\underbrace{(14)(45)(\overline{54})
(41)}_{U_3} \nonumber \\
&= &U_1U_2U_3U_2U^2_3
\end{eqnarray}
where we have again introduced backtracking loops so as to peel off
successively,
the loops corresponding to the different windows.
Note that all these loops are based  at the common  point 1,
because of the introduction of backtracking or \lq thin' loops.
Therefore the  $U_i$'s are free random variables and
\begin{eqnarray}
W(C)&=&<\tr [U_1U_2U_3U_2U^2_3]>=<\tr U_1><\tr [U_2U_3U_2U^2_3]> \nonumber \\
&=&<\tr U_1>(<\tr U_2>^2<\tr [U^3_3]>-
<\tr U_2>^2<\tr [U_3><\tr U^2_3]> \nonumber \\
&&+<\tr [U^2_2]><\tr  U_3><\tr  [U^2_3]>)
\end{eqnarray}
This can be compared with the loop average computed by usual means, once
we express the moments of $U_i$ in terms of the appropriate polynomials,
using ({\ref{Lag}).

\centerline{\epsffile{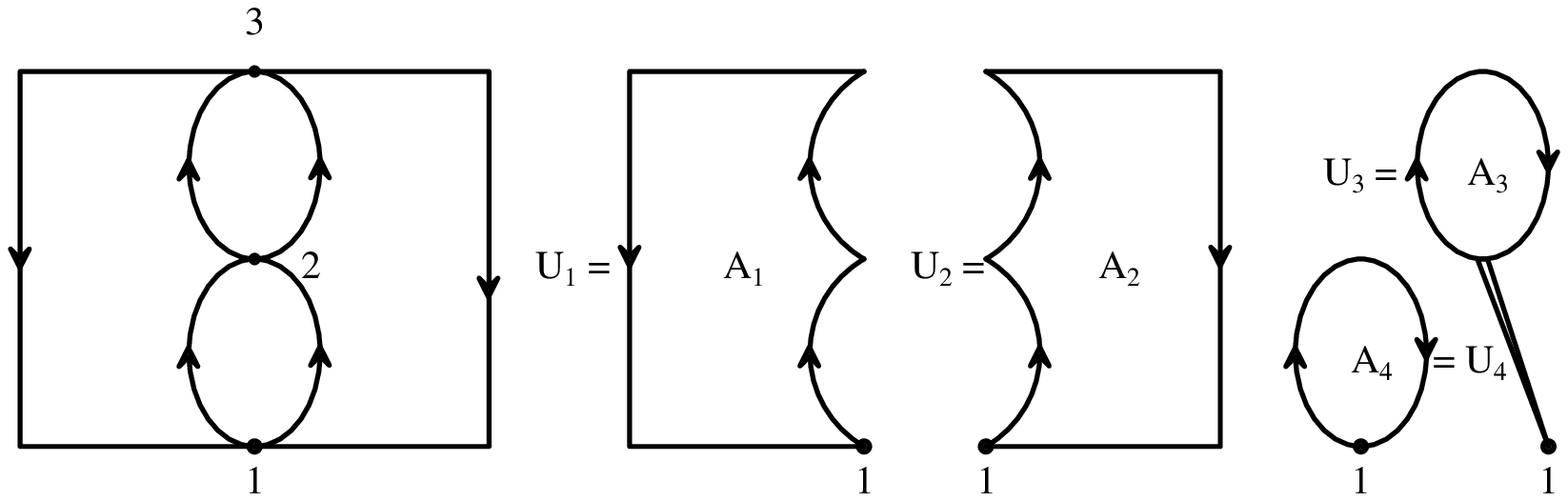}   }
\centerline{ Fig. 8  \hskip .1truein  A Non-Planar Graph Decomposed into Simple
Loops  }
\vskip .2 truein
Finally, consider a case which is \lq non-planar' in the terminology of
\cite{kazkos}, i.e., the loop depicted in Fig.8. With the segments denoted as
shown,
\begin{eqnarray}
 C &= &(12)(23)(31)(\overline{12})(\overline{23})(\overline{31}) \nonumber \\
  & =& (12)(23)(\overline{32})(21)
\underbrace{(12)(\overline{23})(31)}_{U_2}(\overline{12})(\overline{23})(32)
     (\overline{21})\underbrace{(\overline{12})(23)(\overline{31})}_{U_1}
\nonumber \\
&= &
\underbrace{(12)(23)(\overline{32})(21)}_{U_3}U_2\underbrace{(\overline{12})(21)}_{U^{-1}_4}
\underbrace{(12)(\overline{23})(32)(21)}_{U^{-1}_3}
    \underbrace{(12)(\overline{21})}_{U_4}U_1 \\
&= & U_3U_2U^{-1}_4U^{-1}_3U_4U_1
\end{eqnarray}
Once again, the $U_i$'s are free random variables and
therefore the loop average in this case is
\begin{eqnarray}
W(C)&=&<\tr [U_3U_2U^{-1}_4U^{-1}_3U_4U_1]>=<\tr
[U_3U_2U^{-1}_4U^{-1}_3U_4]><\tr U_1> \nonumber \\
&=&<\tr U_1><\tr U_2><\tr [U^{-1}_4U^{-1}_3U_4U_3]>=<\tr U_1><\tr U_2>
\nonumber \\
& &(<\tr U^{-1}_4><\tr U_4>
 -<\tr U^{-1}_4><\tr U^{-1}_3><\tr U_4><\tr U_3> \nonumber \\
 && +<\tr U^{-1}_3><\tr U_3>)
=e^{-\frac{A_1+A_2}{2}}(e^{-A_4} -e^{-(A_4+A_3)}+e^{-A_3})
\end{eqnarray}
which once again reproduces the usual answer.

By now the general procedure must be apparent -- we decompose the loop
starting with the segments bordering the outside and form loops from each of
the windows
giving a coherent orientation.
It is   posssible to
characterise these words algorithmically in terms of the graphical structure,
but
that is not pertinent to our present purpose.
We also see that this process of associating a word
with a loop and then using the recursion relations (\ref{recur}) makes the
computation of
complicated loop averages rather simple, in fact, mechanical.

\subsection{The master loop operators}

We have decomposed an arbitrary loop with holonomy $U_{\Gamma}$
into a word $\Gamma$ built of simple,non-overlapping loops $C_i$ and holonomies
$U_{C_i}$ such that
\begin{equation}
U_{\Gamma}=\prod_{\Gamma} U^{n_i}_{C_i}
\end{equation}
Since the $U_{C_i}$ are free random variables, we can associate master
loop operators to them, $\hat{U}_{C_i}$, by the general construction
of Section 2. Then the master loop operator $\hat{U}_{\Gamma}$ that
reproduces the loop average is
\begin{equation}
W(C)=\left\langle
\Omega|\left(\hat{U}_{\Gamma}=\prod_{\Gamma}\hat{U}^{n_i}_{C_i}\right)
|\Omega\right\rangle .
\end{equation}

We  shall now construct the loop operators $\hat{U}$ (supressing the contour
labels)
in the form
\begin{equation}\label{loopmas}
\hat{U}=\hat a+\sum_{k=0}^{\infty}\omega_{k+1}\hat a^{\dag k}
\end{equation}
The $\omega_k$ can be determined from the $<\tr[U^n]>$ which we
saw, are given by  (\ref{Lag}). We claim that  with
\begin{equation}
\omega_k =(-1)^{k-1}\frac{k^{k-1}}{k!}A^{k-1}e^{-kA/2}
\end{equation}
we have
\begin{equation}
\left\langle \Omega|\hat{U}^n |\Omega\right\rangle = \frac{1}{n}
L_{(n-1)}^1(nA)
e^{-n\frac{A}{2}}=<\tr[U^n]>.
\end{equation}
We shall demonstrate this directly below. We can also
represent $\hat{U}$ in an explicitly unitary manner, as in
the one plaquette model. However, the manifestly unitary form  is not
particularly elegant and we shall not present it here.

\raiseFig{0.1truein}
\moveFigLeft{0truein}
\rightSubFig{field}{0}{1}
Consider an infinitesimal rectangular
loop (as in Fig.9) of area $\Delta A=\Delta x \Delta t$. The master loop
operator associated with it is  (\ref{loopmas}) for $A \rightarrow 0$.Thus to
lowest order in $\Delta A$, we have
\begin{equation}
\hat{U}(\Delta A)=\hat a+(1-\frac{\Delta A}{2})-\Delta A \hat a^{\dag},
\end{equation}
where $\hat a$ refers to the annihilation operator at the point $x$, $\hat
a(x)$.
We can equivalently represent this, by  perfoming a  similarity transformation,
in the form
\begin{equation}\label{uhat}
\hat{U}(\Delta A)=(1-\frac{\Delta A}{2})+i\sqrt{\Delta A}(\hat a+\hat a^{\dag})
=\exp (i\hat{H}) ; \ \
\hat{H}=\sqrt{\Delta A}(\hat a+\hat a^{\dag}).
\end{equation}
This is the explicitly unitary form for $\hat{U}$ for the infinitesimal
loop. Note that if we naively drop the term linear in $\Delta A$, which arises
from
$-\oh H^2=-\oh{\Delta A}(\hat a+\hat a^{\dag})^2 =-\oh{\Delta A} +\cdots$, as
being of
higher order than the $\sqrt{\Delta A}$, then $\hat{U}$  would  not
reproduce the correct leading behaviour in $\Delta A$ in $<\tr[U^n]>$.

But we also know that the holonomy $U$ around such a loop is,
in say, axial gauge
\begin{equation}\label{hol}
U=(1-iA_0(x,t)\Delta t)(1+iA_0(x+\Delta x,t)\Delta t)
 =(1+i\partial_1A_0 \Delta A)
 =\exp (i\partial_1A_0 \Delta A).
\end{equation}
Comparing (\ref{hol}) and (\ref{uhat}) we have
\begin{equation}
\sqrt{\Delta A}\partial_1A_0 =(\hat a+\hat a^{\dag}).
\end{equation}
This is equivalent to the master field of Section 4. Indeed, if we discretise
the
theory. i.e., smear the fields over plaquettes of size
$\Delta A$, then the action reads as
\begin{equation}
S=\frac{1}{2}\sum_{plaquettes}(\Delta A)\Tr(\partial_1A_0)^2.
\end{equation}
We see that $\sqrt{\Delta A}\partial_1A_0$ are Gaussian free random  variables,
represented by $(\hat a+\hat a^{\dag})$, the result we obtained from the loop
operator. Of course, this should come as no real surprise.

It is somewhat less  trivial to  start  from the master gauge field and  to
calculate
  the master loop operators explicitly  for finite loops.
In $QCD_2$, we can do this rather easily since the $\hat U$'s
 have the special property of not just being free random variables  but of also
being {\em  a multiplicative
free family.} This is a concept analogous to the additive free family
that we discussed  in Sec.2.5.  We shall  briefly explain this concept.

The product of two free random variables with distributions, $\mu_1$ and
$\mu_2$
is again a free random variable with some distribution $\mu_3$ denoted by
$\mu_1 \otimes \mu_2$. A one parameter family
of free random variables,  such that $\mu_{t_1} \otimes
\mu_{t_2}=\mu_{t_1t_2}$,
 will be called a multiplicative free family. (Or equivalently
$\mu_{s_1} \otimes \mu_{s_2}=\mu_{s_1+s_2}$, if we redefine
the parameter $t \rightarrow s=\log t$.)

We claim that $\hat{U}(A)$ are a multiplicative free family with the
area $A$ playing the role of the parameter $s$.
In other words, given  two simple loops $C_1$ and $C_2$, based at a point and
non-overlapping, $\hat{U}_{C_1}(A_1)\hat{U}_{C_2}(A_2)$ has the same
distribution
as $\hat{U}_{C_1\circ C_2}(A_1+A_2)$.
Again, there are many ways to see this. One is
from the fact that the heat kernel action is self reproducing and exponentially
dependent on the area of the plaquette.

\moveFigLeft{0truein}
\rightSubFig{rel}{0}{1}
A more explicit way is to note that
$\hat{U}_{C_1}(A_1)\hat{U}_{C_2}(A_2)$
has the same distribution as does
$\hat{U}_{C_1}(A_1)\hat{U}_{C_2}^{\dag}(A_2)$.
This is evident from Fig. 10, where we see that
since $\hat W$ has the
same distribution as $\hat W^{\dag}(A_2)$
and both are independent of $\hat V$.
But $\hat V \hat W^{\dag}$

\noindent is a simple loop of area  equal to the sum of the two areas.

This fact  alone, actually enables us  to construct the
$\hat{U}$ solely from the knowledge of the master gauge field,
i.e. from the knowledge of an  infinitesimal loop.
To do so we need  a non-commutative analog of
the Mellin Transform in ordinary probability theory, which is
multiplicative for the product of two random variables. It
turns out that one can define such a transform  \cite{Voic}---the S-transform,
such that $S_{\mu_1}S_{\mu_2}=$
$S_{\mu_1 \otimes \mu_2}$. For a multiplicative
free family $S_{s_1}S_{s_2}=$
$S_{s_1+s_2}$ (dropping the $\mu$'s.)  $S(z)$ is therefore
exponential in $s$ in  this case. The function $S(z)$ for a non-commutative
random
variable $U$ is constructed as
follows :
If
\begin{equation}
\phi(j)=\sum_{n=1}^{\infty}\vev{\hat U^n} j^n,
\end{equation}
then construct the inverse function  $\chi(z)$, i.e. $\phi(\chi(z))=z$.
The S-transform is defined as:
\begin{equation}
S(z)=\frac{1+z}{z}\chi(z).
\end{equation}

Since the $\hat{U}$'s are multiplicative, we can use the S-transform
of an infinitesimal loop to obtain the exact S-transform
for one of finite area, knowing that it must necessarily exponentiate.
For the infinitesimal rectangular loop in Fig.9.
we saw, in axial gauge, that
\begin{equation}
U=(1+i\partial_1A_0 \Delta A)=\exp (i\partial_1A_0 \Delta A).
\end{equation}
Arguing backwards now, from the discretised action for $QCD_2$,
for which $\sqrt{\Delta A}\partial_1A_0 =(\hat a+\hat a^{\dag})$, we have
\begin{equation}
\hat{U}= \exp (i\hat{H}) \ \ \ \ \hat{H}=\sqrt{\Delta A}(\hat a+\hat a^{\dag}).
\end{equation}
Now,
\begin{equation}
\phi(j)=\sum_{n=1}^{\infty}\vev{\hat U^n} j^n=\sum_{n=1}^{\infty}(1-n^2\Delta
A/2)j^n ,
\end{equation}
keeping only terms of order $\Delta A$.
The sum can be performed giving
\begin{equation}
\phi(j)=\frac{j}{1-j}-\frac{\Delta A}{2}j \frac{(3-j)}{(1-j)^3}.
\end{equation}
Equating this to $z$ and solving for $j \equiv \chi(z)$ (again to lowest order
in $\Delta A$ only) gives
\begin{equation}
\chi_{\Delta A}(z)=\frac{z}{1+z}(1+\Delta A(1+2z)).
\end{equation}
Therefore $S_{\Delta A}(z)$, defined as $\chi_{\Delta A}(z)\frac{1+z}{z}$, is
equal to
$(1+\Delta A(1+2z))$
for an infinitesimal loop.
For finite area, this exponentiates as expected, to give
\begin{equation}
S_A(z)=e^{\frac{A}{2}(1+2z)} \Rightarrow
\chi_A(z)=\frac{z}{1+z}e^{\frac{A}{2}(1+2z)}.
\end{equation}

We can now  use the $S_A(z)$, which we obtained from the master
gauge field, to give an alternative derivation of (\ref{Lag}) for
$<\tr [U^n]>$.
Since
\begin{eqnarray}
\phi(e^{-i\theta})&=&\sum_{n=1}^{\infty}<\tr[U^n]>e^{-in\theta} \nonumber \\
 \Rightarrow <\tr[ U^n]>&=&\frac{1}{2\pi}\int
\phi(e^{-i\theta})e^{in\theta}d\theta,
\end{eqnarray}
we have, with $\chi(z)=e^{-i\theta}$,
\begin{equation}
<\tr [U^n]>=\oint \frac{dz}{2\pi i}z[\chi(z)]^{-(n+1)}\chi^{\prime}(z),
\end{equation}
or on integrating by parts
\begin{equation}
<\tr[ U^n]>=\frac{1}{n}\oint \frac{dz}{2\pi i}[\chi(z)]^{-n}.
\end{equation}
In the case at hand $\chi_A(z)=\frac{z}{1+z}e^{\frac{A}{2}(1+2z)}$
and hence
\begin{equation}
<\tr [U(A)^n]>=\frac{1}{n}\oint \frac{dz}{2\pi i}
(1+\frac{1}{z})^n e^{-n\frac{A}{2} (1+2z)}
\end{equation}
which is (\ref{Lag}).

We can now  see why the Hopf equation arises in the QCD$_2$. In fact it, or its
generalization
\re{genhop},
 will appear
for any multiplicatively free family of random variables.
Define the resolvent of $U(A)$ to be
\begin{equation}
R(\zeta, A)=\sum_{n=0}^{\infty}<\tr [U(A)^n]>\zeta^{-(n+1)}.
\end{equation}
By definition $\phi$ is related to the resolvent as
\begin{eqnarray}\label{rphi}
R(\zeta, A)=\frac{1}{\zeta}(\phi(\frac{1}{\zeta})+1)
\Rightarrow R(\frac{1}{\chi(z)}, A)= \chi(z)(z+1).
\end{eqnarray}
Now $ \exp[i\theta]= 1/\chi(z)=(1+z)/z\exp[-A/2(1+2z)]$,
from which it follows that
\begin{equation}
\theta(z)= iA (z+\oh) - i \log{1+z\over z}.
\end{equation}
Therefore, using \re{rphi} and the fact that $\phi(\chi(z))=z$, we have
\begin{equation}
R(e^{i\theta}= {1\over \chi(z)}, A) = e^{-i\theta}(\phi(e^{-i\theta}=\chi(z))
+1)= e^{-i\theta}(z+1)
\Rightarrow e^{i\theta}R(e^{i\theta}, A)= 1 +z.
\end{equation}
Redefining $w= i(z +\oh)$ we have
\begin{equation}
\theta(w)=Aw-i \log{w-{i\over 2} \over w+{i\over 2}}.
\end{equation}
Then $F( \theta, A) \equiv i[ e^{i\theta}R(e^{i\theta}, A)- \oh] =w$ satisfies
\begin{equation}
{ \partial F \over \partial A} + F{ \partial F \over \partial \theta} = 0.
\end{equation}

We can also employ $S_A(z)$ to explicitly compute
$\hat{U}$  in the form given by  (\ref{loopmas}). Thus if
\begin{equation}
U(z)=\frac{1}{z} + \sum_{k=0}^{\infty}\omega_{k+1}z^k
\end{equation}
then $U(z)$ is the inverse of the resolvent
and
\begin{eqnarray}
R(\frac{1}{\chi(z)})= \chi(z)(z+1)
\Rightarrow U((z+1)\chi(z)) =\frac{1}{\chi(z)}.
\end{eqnarray}
Therefore
\begin{equation}
U(ze^{\frac{A}{2}(1+2z)})=\frac{1+z}{z}e^{-\frac{A}{2}(1+2z)}.
\end{equation}
Consequently,  if $z(y)$ is determined from
\begin{equation}\label{zy}
ze^{\frac{A}{2}(1+2z)}=y,
\end{equation}
then
\begin{equation}
U(y)=\frac{1}{y}(1+z(y)).
\end{equation}
This is essentially the result we need. We have obtained $U(y)$,
albeit
thus far in an  implicit form. To obtain the coefficients $\omega_k$, we must
examine the relation (\ref{zy}) more carefully.
We have
\begin{eqnarray}\label{yU}
yU(y)&=&1+z(y)=1+\sum_{k=1}^{\infty}\omega_k y^k\nonumber  \\
&=&1+\sum_{k=1}^{\infty}
\omega_k z^k e^{k\frac{A}{2}(1+2z)} \Rightarrow
z=\sum_{k=1}^{\infty}\omega_k z^k e^{k\frac{A}{2}(1+2z)}.
\end{eqnarray}
This is what determines the coefficients $\omega_k $. In fact, if we redefine
$\omega_k =A^{k-1}e^{-k\frac{A}{2}}c_k$ and $z\rightarrow Az$,
then  (\ref{yU})
becomes an equation for the $c_k$'s.
\begin{equation}
z=\sum_{k=1}^{\infty}c_k z^k e^{kz}.
\end{equation}
All the area dependence is gone and the $c_k$ are just some numbers determined
recursively by the above equation. In fact, the recursion relation is non
trivial
\begin{equation}
c_k=-\sum_{r=1}^{k-1}c_{k-r}\frac{(k-r)^r}{r!}
\end{equation}
It can be checked that the $c_k$ are precisely
$(-1)^{k-1}\frac{k^{k-1}}{k!}$
due to the non-trivial combinatorial relation
\begin{equation}
(-1)^{k-1}\frac{k^{k-1}}{k!}=(-1)^{k-r}\frac{(k-r)^{k-1}}{(k-r)!r!}
\end{equation}
But we can actually, rather simply argue that $\omega_k$ have to take this
form.
That is, it is simply the highest power of $A$ term that appears in $<\tr
U^k>$.
This follows since $<\tr U^k>$ is, a polynomial with highest power of $A$
being $A^{k-1}$ multiplying, of course, the $e^{-kA/2}$.
Since
$\omega_n \propto A^{n-1}$, the only term in $<0|{\hat U}^k|0>$
that can contribute
an $A^{k-1}$ term  is $\omega_k $ (other polynomial
terms are of the form $\omega_{i_1}\omega_{i_2} \ldots \omega_{i_r}$
with $\sum_r i_r =k$ and thus are always of lower
order). Therefore $\omega_k $ must be precisely the term appearing
with $A^{k-1}$ in $<\tr U^k>$ which is exactly what was given earlier
(as can be checked from the expansion of the  $L^1_{(n-1)}$).

\section{Conclusions}

In this paper we have  reviewed the basic  concepts of non-commutative
probability theory
and applied them to the large $N$ limit of matrix models. We discussed at
length the work
of Voiculescu on the properties and representation of free random variables.
Since independent matrix models at $N = \infty$ are free random variables this
appears to
be the appropriate framework for constructing the master field.
We discussed  some of these models,
including the one-plaquette model
where we explicitly constructed the master field.
We also discussed, at length, QCD$_2$. In an axial gauge this theory can be
regarded
as a theory of independent matrices
and thus we could give an explicit construction of the master gauge field.
We also showed that there exists a gauge in which the master gauge
field is spacetime independent. We also constructed
master loop operators based on the observation that simple loops
corresponded to free random variables and that any loop could be decomposed
into words built out of simple loops. The simple structure of
QCD$_2$ is then a consequence of fact that these form a multiplicative free
family.

The most suprising and exciting of our results, however, is
the extension of these techniques to deal with the
general matrix model, in which the matrices
do not have independent distributions and are coupled.
Based on our observation that the generating function, introduced by Voiculescu
to
construct the representation of
an independent random variable,
can be identified as the generating function of connected planar Green's
functions,
we were able to construct the master field for any and all matrix models.
Remarkably the Hilbert space in which the master fields are represented is
unchanged---it is the   Fock space
generated by a collection of creation and annihilation
operators satisfying the Cuntz algebra---one for each matrix variable.

{}From some points of view our construction is somewhat dissapointing. First,
although we
have an explicit construction of the master field for any matrix model
in a well defined Hilbert space, to actually write the master field explicitly
would require a knowledge of all the connected Green's functions, which is
tantamount to
solving the theory. Thus from this point of view all we have done is to
repackage the
unknown solution. Second, we have almost as many degrees of freedom as before.
The Hilbert
space in which the master field is represented is almost as big as the full
Hilbert space of
the quantum field theory--i.e., there is an independent creation operator
for each independent field variable. The only reduction is by a factor of
$N^2$,
since the large $N$ limit has been taken,

However, we believe that this reformulation is valuable. Clearly this
is the appropriate framework for formulating the $N=\infty$ theory.
It also suggests new approaches towards solving the theory by constructing the
master field---now a well defined operator in a well defined space.
For example, one approach might be to explore the operator equations of motion
for the
master field, as we have discussed above. Do these,  for example, follow from
some kind
of variational principle that could be the basis for an approximation scheme?
Can one develop similar techniques for the Hamiltonian formulation of
large $N$ theories?

\vskip .5truein
{\Large{\bf Acknowledgements}}

We would like to thank  Mike Douglas,  Andrei Matytsin, and Sasha Migdal  for
 discussions. D.G thanks I. Singer for discussions and for bringing the work of
Voiculescu to his attention.

\end{document}